\input harvmac
\input epsf


\lref\Heine{
 H. Heine, "Kugelfunktionen", Berlin, 1878 and 1881. Reprinted Physica
Verlag, W\"urzburg, 1961;~
 G. Szeg\"o, "Orthogonal Polynomials", AMS, New York, 1959.
}

\lref\MukhopadhyayEN{
P.~Mukhopadhyay and A.~Sen,
``Decay of unstable D-branes with electric field,''
JHEP {\bf 0211}, 047 (2002)
[arXiv:hep-th/0208142].
}

\lref\MaloneyCK{
A.~Maloney, A.~Strominger and X.~Yin,
``S-brane thermodynamics,''
JHEP {\bf 0310}, 048 (2003)
[arXiv:hep-th/0302146].
}

\lref\JonesRG{
G.~Jones, A.~Maloney and A.~Strominger,
``Non-singular solutions for S-branes,''
arXiv:hep-th/0403050.
}

\lref\TasinatoDY{
G.~Tasinato, I.~Zavala, C.~P.~Burgess and F.~Quevedo,
``Regular S-Brane Backgrounds,''
arXiv:hep-th/0403156.
}

\lref\Gaiotto{
D.~Gaiotto, N.~Itzhaki and L.~Rastelli,
``Closed strings as imaginary D-branes,''
Nucl.\ Phys.\ B {\bf 688}, 70 (2004)
[arXiv:hep-th/0304192].}

\lref\Bergman{
O.~Bergman and S.~S.~Razamat,
``Imaginary time D-branes to all orders,''
arXiv:hep-th/0402124.}
\lref\StromDS{
A.~Strominger,
``The dS/CFT correspondence,''
JHEP {\bf 0110}, 034 (2001)
[arXiv:hep-th/0106113];~~~
``Inflation and the dS/CFT correspondence,''
JHEP {\bf 0111}, 049 (2001)
[arXiv:hep-th/0110087].
}

\lref\VijayJanDjordjeDS{
V.~Balasubramanian, J.~de Boer and D.~Minic,
``Mass, entropy and holography in asymptotically de Sitter spaces,''
Phys.\ Rev.\ D {\bf 65}, 123508 (2002)
[arXiv:hep-th/0110108];~
``Holography, time and quantum mechanics,''
arXiv:gr-qc/0211003;~
``Exploring De Sitter Space And Holography,''
Class.\ Quant.\ Grav.\  {\bf 19}, 5655 (2002)
[Annals Phys.\  {\bf 303}, 59 (2003)]
[arXiv:hep-th/0207245].
}

\lref\LarsenWC{
F.~Larsen, A.~Naqvi and S.~Terashima,
``Rolling tachyons and decaying branes,''
JHEP {\bf 0302}, 039 (2003)
[arXiv:hep-th/0212248].
}

\lref\StromingerPC{
A.~Strominger,
``Open string creation by S-branes,''
arXiv:hep-th/0209090.
}

\lref\SenIN{A.~Sen,
``Rolling tachyon,''
JHEP {\bf 0204}, 048 (2002)
[arXiv:hep-th/0203211];~~~
A.~Sen,
``Tachyon matter,''
JHEP {\bf 0207}, 065 (2002)
[arXiv:hep-th/0203265];~~~
A.~Sen,
``Field theory of tachyon matter,''
Mod.\ Phys.\ Lett.\ A {\bf 17}, 1797 (2002)
[arXiv:hep-th/0204143].
}

\lref\UVfinite{
J.~L.~Karczmarek, H.~Liu, J.~Maldacena and A.~Strominger,
``UV finite brane decay,''
JHEP {\bf 0311}, 042 (2003)
[arXiv:hep-th/0306132].
}

\lref\Wang{
J.~E.~Wang,
``Twisting S-branes,''
arXiv:hep-th/0403094.
}

\lref\LuIV{
H.~Lu, C.~N.~Pope and J.~F.~Vazquez-Poritz,
``From AdS black holes to supersymmetric flux-branes,''
arXiv:hep-th/0307001;~~
H.~Lu and J.~F.~Vazquez-Poritz,
``Four-dimensional Einstein Yang-Mills de Sitter gravity from eleven
dimensions,''
Phys.\ Lett.\ B {\bf 597}, 394 (2004)
[arXiv:hep-th/0308104];~~
H.~Lu and J.~F.~Vazquez-Poritz,
``Non-singular twisted S-branes from rotating branes,''
JHEP {\bf 0407}, 050 (2004)
[arXiv:hep-th/0403248].
}

\lref\Bateman{A. Erdelyi, ed., ``Higher Transcendental Functions,
Vol. 1-3", Kreiger (Malabar, Florida) 1981.}

\lref\Mathworld{Eric W. Weisstein et al. ``Hurwitz Zeta Function." From MathWorld--A Wolfram Web Resource. http://mathworld.wolfram.com/HurwitzZetaFunction.html;  ~~~ Also see http://functions.wolfram.com/ZetaFunctionsandPolylogarithms/Zeta2/06/02/.}

\lref\Douglas{
M.~R.~Douglas,
``Conformal field theory techniques for large N group theory,''
arXiv:hep-th/9303159.
}

\lref\ConstableRC{
N.~R.~Constable and F.~Larsen,
``The rolling tachyon as a matrix model,''
JHEP {\bf 0306}, 017 (2003)
[arXiv:hep-th/0305177].
}

\lref\GutperleAI{
M.~Gutperle and A.~Strominger,
``Spacelike branes,''
JHEP {\bf 0204}, 018 (2002)
[arXiv:hep-th/0202210].
}
\lref\GutperleXF{
M.~Gutperle and A.~Strominger,
``Timelike boundary Liouville theory,''
Phys.\ Rev.\ D {\bf 67}, 126002 (2003)
[arXiv:hep-th/0301038].
}

\lref\snaith{
J.P.~Keating and N.C.~Snaith,
``Random Matrix Theory and $\zeta(1/2 + i t)$'',
Commun.\ Math.\ Phys. {\bf 214}, 57 (2000).
}

\lref\BorodinOkounkov{
A.~Borodin and A.~Okounkov,
``A Fredholm determinant formula for Toeplitz determinants,"
Int.\ Eqs.\ Oper.\ Th.\ 37 (2000) 386-396
[arXiv:math.CA/9907165].
}

\lref\BasorWidom{ E.~L.~Basor and H.~Widom, ``On a Toeplitz
determinant identity of Borodin and Okounkov," Int.\ Eqs.\ Oper.\
Th.\ 37 (2000) 397-401 [arXiv:math.FA/9909010]. }

\lref\DouglasUP{
M.~R.~Douglas, I.~R.~Klebanov, D.~Kutasov, J.~Maldacena, E.~Martinec and N.~Seiberg,
``A new hat for the c = 1 matrix model,''
arXiv:hep-th/0307195.
}

\lref\BakerForresterPearce{
T.~H.~Baker, P.~J.~Forrester and P.~A.~Pearce,
``Random matrix ensembles with an effective extensive
external charge",
[arXiv:cond-mat/9803355].
}

\lref\GradshteynRyzhik{
I.~S.~Gradshteyn and I.~M.~Ryzhik,
``Table of Integrals, Series, and Products'',
corrected and enlarged edition, Academic Press 1980.
}

\lref\OkudaYD{ T.~Okuda and S.~Sugimoto,
``Coupling of rolling tachyon to closed strings,''
Nucl.\ Phys.\ B {\bf 647}, 101 (2002) [arXiv:hep-th/0208196].
}

\lref\deBoerHD{ J.~de Boer, A.~Sinkovics, E.~Verlinde and J.~T.~Yee,
``String interactions in c = 1 matrix model,''
arXiv:hep-th/0312135.
}

\lref\LambertZR{ N.~Lambert, H.~Liu and J.~Maldacena,  ``Closed
strings from decaying D-branes,'' arXiv:hep-th/0303139.
}


\lref\SenVV{
A.~Sen,
``Time evolution in open string theory,''
JHEP {\bf 0210}, 003 (2002)
[arXiv:hep-th/0207105].
}

\lref\GutperleXF{
M.~Gutperle and A.~Strominger,
``Timelike boundary Liouville theory,''
Phys.\ Rev.\ D {\bf 67}, 126002 (2003)
[arXiv:hep-th/0301038].
}

\lref\MaloneyCK{
A.~Maloney, A.~Strominger and X.~Yin,
``S-brane thermodynamics,''
JHEP {\bf 0310}, 048 (2003)
[arXiv:hep-th/0302146].
}

\lref\GaiottoRM{
D.~Gaiotto, N.~Itzhaki and L.~Rastelli,
``Closed strings as imaginary D-branes,''
arXiv:hep-th/0304192.
}

\lref\SenBC{
A.~Sen,
``Open and closed strings from unstable D-branes,''
Phys.\ Rev.\ D {\bf 68}, 106003 (2003)
[arXiv:hep-th/0305011].
}

\lref\SenXS{
A.~Sen,
``Open-closed duality at tree level,''
Phys.\ Rev.\ Lett.\  {\bf 91}, 181601 (2003)
[arXiv:hep-th/0306137].
}

\lref\OkuyamaJK{
K.~Okuyama,
``Comments on half S-branes,''
JHEP {\bf 0309}, 053 (2003)
[arXiv:hep-th/0308172].
}

\lref\SenNU{
A.~Sen,
``Rolling tachyon,''
JHEP {\bf 0204}, 048 (2002)
[arXiv:hep-th/0203211].
}

\lref\SenIV{
A.~Sen,
``Open-closed duality: Lessons from matrix model,''
arXiv:hep-th/0308068.
}

\lref\SenZM{
A.~Sen,
 ``Rolling tachyon boundary state, conserved charges and two dimensional string theory,''
arXiv:hep-th/0402157.
}

\lref\selbergs{
M.~Mehta,
``Random Matrices", Academic Press, London, second edition, 1991. }

\lref\FredenhagenUT{
S.~Fredenhagen and V.~Schomerus,
``On minisuperspace models of S-branes,''
JHEP {\bf 0312}, 003 (2003)
[arXiv:hep-th/0308205].
}

\lref\BergmanPB{
O.~Bergman and S.~S.~Razamat,
``Imaginary time D-branes to all orders,''
arXiv:hep-th/0402124.
}

\lref\KarczmarekXM{
J.~L.~Karczmarek, H.~Liu, J.~Maldacena and A.~Strominger,
``UV finite brane decay,''
JHEP {\bf 0311}, 042 (2003)
[arXiv:hep-th/0306132].
}

\lref\JonesRG{
G.~Jones, A.~Maloney and A.~Strominger,
``Non-singular solutions for S-branes,''
arXiv:hep-th/0403050.
}

\lref\SchomerusVV{ V.~Schomerus, ``Rolling tachyons from Liouville
theory,'' JHEP {\bf 0311}, 043 (2003) [arXiv:hep-th/0306026].
}

\lref\TeschnerRV{
J.~Teschner,
``Liouville theory revisited,''
Class.\ Quant.\ Grav.\  {\bf 18}, R153 (2001)
[arXiv:hep-th/0104158].
}

\lref\StromingerFN{
A.~Strominger and T.~Takayanagi,
``Correlators in timelike bulk Liouville theory,''
Adv.\ Theor.\ Math.\ Phys.\  {\bf 7}, 369 (2003)
[arXiv:hep-th/0303221].
}

\lref\Volkertoappear{
S.~Fredenhagen and V.~Schomerus, ``Exact Rolling Tachyons", to appear.}

\newcount\figno
\figno=0
\def\fig#1#2#3{
\par\begingroup\parindent=0pt\leftskip=1cm\rightskip=1cm\parindent=0pt
\baselineskip=11pt
\global\advance\figno by 1
\midinsert
\epsfxsize=#3
\centerline{\epsfbox{#2}}
\vskip 12pt
{\bf Fig.\ \the\figno: } #1\par
\endinsert\endgroup\par
}
\def\figlabel#1{\xdef#1{\the\figno}}
\def\encadremath#1{\vbox{\hrule\hbox{\vrule\kern8pt\vbox{\kern8pt
\hbox{$\displaystyle #1$}\kern8pt}
\kern8pt\vrule}\hrule}}
\baselineskip12pt

\def\ap{\alpha'}
\def\a{\alpha}
\def\p{\partial}

\def\wb{\bar{w}}
\def\zb{\bar{z}}

\def\Ebf{{\bf E}}

\def\CH{{\cal H}}

\Title{\vbox{\baselineskip12pt \hbox{hep-th/0404039}
\hbox{UPR-T-1074, HIP-2004-TH/12} \hbox{UCLA-04-TEP-20, ITFA-2004-11}}}
{\vbox{\centerline { String Scattering from Decaying Branes}} }
\centerline{{\bf Vijay Balasubramanian\foot{vijay@physics.upenn.edu},
Esko Keski-Vakkuri\foot{esko.keski-vakkuri@helsinki.fi}, Per
Kraus\foot{pkraus@physics.ucla.edu} and Asad
Naqvi\foot{anaqvi@science.uva.nl}} }

\medskip
\centerline{${}^1$David Rittenhouse Laboratory, University of
Pennsylvania, Philadelphia, PA 19104, USA}
\centerline{${}^2$Helsinki Institute of Physics and Department of
Physical Sciences,} \centerline{P.O. Box 64, FIN-00014 University
of Helsinki, Finland} \centerline{${}^3$Department of Physics,
University of California, Los Angeles, 90095, USA}
\centerline{${}^4$Institute for Theoretical Physics, University of
Amsterdam, The Netherlands}

\vskip .1in

\centerline{\bf Abstract} {We develop the general formalism of
string scattering from decaying D-branes in bosonic string theory.
In worldsheet perturbation theory, amplitudes can be written as a
sum of correlators in a grand canonical ensemble of unitary random matrix
models, with  time setting the fugacity.  An approach employed in
the past for computing amplitudes in this theory involves an
unjustified analytic continuation from special integer momenta.
  We give an
alternative formulation which is well-defined for general momenta.
We study the emission of closed strings from a decaying D-brane with
initial conditions perturbed by the addition of an open string
vertex operator.    Using an integral formula due to Selberg, the
relevant amplitude is expressed in closed form in terms of zeta
functions.   Perturbing the initial state can suppress or enhance the emission
of high energy closed strings for extended branes, but enhances it
for D0-branes.    The closed string two point function is
expressed as a
 sum of Toeplitz determinants of certain hypergeometric functions.  A large N limit theorem due to
  Szeg\"{o}, and its extension due to Borodin and Okounkov, permits us to compute approximate results
   showing that previous naive analytic continuations amount to the large N approximation of the
   full result.
 We also give a free fermion formulation of scattering from decaying D-branes and  describe the relation
 to a grand canonical ensemble
for a 2d Coulomb gas. }

\noindent
\Date{}
\newsec{Introduction}

D-branes, as solitons of open string theory that are localized in
space, have given many insights into nonperturbative phenomena in
string theory such as string duality, and resolve many timelike
singularities of General Relativity including those inside some
black holes.    In addition, they  lead to the holographic
description  of 11 dimensional asymptotically flat space via a
Matrix model and of spaces with a negative cosmological constant
in terms of a dual conformal field theory.   Many of these
developments, and particularly the last, arose from an
understanding of D-brane dynamics -- namely how closed strings
scatter from such solitons which are quantized in terms of open
string fluctuations.

It is of great interest to understand similar issues in the time
dependent context of rapidly expanding universes, particularly in
view of the likely occurrence of a Big Bang followed by inflation
in the early universe and the possible presence of a positive
cosmological constant  $\Lambda >0$ now. Exploration of the
symmetries and structure of universes with positive $\Lambda$ has
suggested that if they have a holographic description the dual
might be related to a Euclidean CFT living on the early or late
time boundaries of such spacetimes (\StromDS, \VijayJanDjordjeDS).\foot{One heuristic way of
motivating  this is to note that  Lorentzian de Sitter space and
Euclidean AdS space are solutions to $-X_0^2 + X_1^2 + \cdots
X_d^2 = \pm l^2$ embedded in flat $(d+1)$ dimensional space. Both
hyperboloids are thus different  real sections of the same complex
manifold.} Time, in such a picture, emerges holographically via
the RG flow of the Euclidean field theory dual in analogy with the
emergence of the radial direction of AdS space from the RG flow of
a Lorentzian field theory (\StromDS, \VijayJanDjordjeDS).   In order for such a picture to be
actually realized in string theory one would need some kind of
D-brane localized in time, called an S-brane in \GutperleAI, with
a decoupling limit relating the closed strings on the spacetime
background to the open strings quantizing the brane. One might
hope that the spacetime near such a Euclidean brane would be
rapidly expanding by analogy with the rapid expansion of the
transverse space in the vicinity of a conventional brane.  In the
decoupling limit for standard D-branes it is precisely this rapid
transverse expansion that gives rise to the exponential increase
in the volume of AdS spaces as the boundary is approached.   Thus
one might hope that a Euclidean brane would lead to a spacetime
exponentially expanding in time in a suitable decoupling limit.

Sen has proposed \SenIN\ that S-branes are concretely realized in
string theory by the exact boundary CFTs representing decaying
branes. In bosonic string theory the exactly marginal boundary
interaction \eqn\bndpert{ S_{{\rm bndy}} = \lambda \, \int \! dt
\, e^{X^0} } (where $X^0$ is the timelike scalar) describes such a
brane \StromingerPC, \LarsenWC, \GutperleXF.\foot{There are also
$\cosh X^0$ and $\sinh X^0$ interactions,   which describe brane
formation and subsequent decay \SenIN.   Note the analogy with the
inflationary and global parametrizations of de Sitter space,
namely $ds^2 = -dt^2 + e^{t} d\vec{x}^2$ versus $ds^2 = -dt^2 +
\cosh(t)^2 d\Omega^2$.} The basic results that have been obtained
so far are:

\lref\McGreevyKB{ J.~McGreevy and H.~Verlinde, ``Strings from
tachyons: The c = 1 matrix reloaded,'' JHEP {\bf 0312}, 054 (2003)
[arXiv:hep-th/0304224].
}

\lref\KlebanovKM{ I.~R.~Klebanov, J.~Maldacena and N.~Seiberg,
``D-brane decay in two-dimensional string theory,'' JHEP {\bf
0307}, 045 (2003) [arXiv:hep-th/0305159].
}

\item{1.} At vanishing $g_s$ the brane decays to ``tachyon
matter'' with  energy but vanishing pressure \SenIN. 
\item{2.} At
finite coupling there is tremendous production of very heavy
non-relativistic closed strings \LambertZR. 
\item{3.} It has been
proposed that the resulting coherent state of heavy closed strings
is an equivalent description of the tachyon matter. 
\item{4.} In
view of (2) and (3) it is proposed that there is a new open-closed
duality hinting at a new kind of holography \SenBC, \SenXS.
\item{5.} There is a close relation with the dynamics of decaying
unstable branes in the $c=1$ matrix model that suggests that the
picture (1) - (4) is essentially correct (\McGreevyKB, \KlebanovKM,
\DouglasUP,\SenIV).

To understand the structure of tachyon matter, and to explore the
possibility of a decoupling limit leading to timelike holography,
a central problem is to compute general scattering amplitudes of
closed and open strings from the decaying brane.  Closed strings
can be used particularly to probe the structure of the tachyon
matter final state that exists after the brane decays, and open
strings, since they will only be present in the spectrum at early
times, can be used to change the initial brane configuration which
is decaying.

\lref\GoulianQR{ M.~Goulian and M.~Li, ``Correlation Functions In
Liouville Theory,'' Phys.\ Rev.\ Lett.\  {\bf 66}, 2051 (1991).
}

Thus, in Sec.~2 and 3 we develop the general formalism of string scattering from decaying branes in
bosonic string theory.  In a perturbative approach, amplitudes can be written as a sum of
correlators in a grand canonical ensemble of unitary matrix models, with  time setting
the fugacity.   If we integrate over the zero-mode time coordinate at the outset, the amplitudes are
connected by analytic continuation to a single unitary matrix model, the rank
of which is related to the
total energy carried by the external vertex operators.  Based on
the approach which was employed with success in bulk
Liouville theory \GoulianQR, it was proposed in
(\GutperleXF, \ConstableRC) to compute  bulk correlation
functions by ``analytically continuing'' the vertex operator momenta
from discrete integer values where direct computation is simple.  This procedure
has long been known to be somewhat
questionable since an analytic function cannot be defined from just the data at a discrete set of
points without additional constraints such as knowledge of the behavior  at infinity or
consistency conditions imposed by symmetries.   In the case of Liouville theory, suitable consistency conditions were found (see the review \TeschnerRV) and one can try to exploit the resulting techniques  to explore  the physics of decaying branes (\GutperleXF, \StromingerFN, \SchomerusVV, \FredenhagenUT, \Volkertoappear).   For full branes, an interesting prescription to compute $n$-point closed string disk amplitudes was given in terms of viewing the brane as an array of ordinary D-branes in imaginary time \Gaiotto (for generalization to higher genus,  see \Bergman). In this paper,  we describe an alternative
formulation of the scattering amplitude calculation which is well-defined in an open subset of
the complex energy plane and therefore permits reliable analytic
continuation.  We will study in detail the open-closed and
closed-closed two point functions, and compare our results to
previous work. 

 In Sec.~4, we study the emission of closed strings from a decaying brane with initial conditions
 perturbed by the addition of an open string vertex operator.\foot{Another approach to these amplitudes will be described in \Volkertoappear.}     Using an integral formula due to
 Selberg, the relevant amplitude is expressed in closed form in terms of zeta functions.   Perturbing
 the initial state in this way can either suppress or enhance the emission of high energy closed strings for extended  branes, but always enhances it for D0-branes.       This is consistent with the picture that D0-branes decay entirely into closed strings, with very heavy closed strings making up the tachyon matter state, while higher dimensional branes can decay partly into unstable lower dimensional branes.  However, the enhancement of emission that we find in some cases will increase the direct  closed string emission even by the higher dimensional branes.  We discuss the consequences for the hypothesis that tachyon matter is nothing but a state of  very heavy closed strings.   In Sec.~5, the closed string two point function is expressed as a sum on $N$ of Toeplitz determinants of certain $N\times N$ matrices of hypergeometric functions.  A large-N limit theorem due to Szeg\"{o}, and its extension due to Borodin and Okounkov, permit us to compute approximate results that show that  previously used methods to compute scattering amplitudes in the decaying brane amount to a leading  large $N$ approximation from our perspective.   In Appendix C we give a free fermion formulation of scattering from decaying branes by extending old techniques of Douglas \Douglas.  In Appendix D we describe the relation of the decaying brane correlators to a grand canonical  ensemble for a classical Coulomb gas.

\newsec{Review of closed string tachyon scattering from D-branes}

To establish notation and conventions, in this section we review
the standard computation of closed string tachyon scattering from
a D-brane.  This amounts to computing the bulk two point function on the
disk, with Neumann and Dirichlet boundary conditions in the various
directions.

\subsec{Kinematics}

\lref\Polchinski{J. Polchinski, ``String Theory'', Cambridge University
Press, 1998}

We follow the conventions of Polchinski \Polchinski, with
 $\eta^{\mu\nu}= (-,+,+,+, \ldots)$, and $\ap=1$.
The closed string tachyon vertex operator is
\eqn\secba{e^{i k \cdot X}, \quad k^2 = -m^2 = 4 .}
For scattering from a Dp-brane, we divide the momenta into the
$p+1$ parallel components $k^\parallel$, and the $25-p$ transverse
components $k^\perp$. We write the momenta of the two closed
string tachyons as $k_1 = (k^\parallel, k_1^\perp)$ and $k_2 =
(-k^\parallel, k_2^\perp)$. The Mandelstam variables are defined
as
\eqn\secbc{ s = 2 (k^\parallel)^2, \quad t= k_1 \cdot k_2.}

We can factorize the amplitude in the closed or open string channels, and find
poles at the location of physical string states:
\eqn\secbd{\eqalign{ {\rm closed:}~~ k^2 & = - m^2 =-4 (N-1),\cr {\rm open:}~~ k^2
&= - m^2 =- (N-1)}}
with $N=0,1,2,\ldots$.  This implies that poles occur at
\eqn\secbe{\eqalign{ {\rm closed:}~~  t& = -2(N+1) \cr {\rm
open :}~~  s& = -2(N-1).}}

\subsec{Correlators on disk}

We work on the unit disk, $|z| \leq 1$.  As is standard, we separate
out the zero modes from the  worldsheet fields $X^\mu(z,\zb)$ by writing
\eqn\ba{X^\mu(z,\zb) = x^\mu + X^{' \mu}(z,\zb)}
with $ \int \! d^2 z \, X^{'\mu}(z,\zb)=0$.  The zero mode
integrals are done at the end of the calculation, and enforce
momentum conservation in Neumann directions.

The Neumann and Dirichlet correlators on the disk are\foot{In fact,
there is a subtlety in defining the Neumann correlator.  The Green's
function which obeys Neumann boundary conditions and vanishes when integrated
over the disk is $- {1 \over 2} \eta^{\mu\nu}\left[ \ln |z-w|^2
+ \ln|1 - z \wb|^2 -z\zb - w\wb -c \right]$, where $c$ is a number which
depends on the choice of worldsheet metric.  The extra terms drop out
after using spacetime momentum conservation (see section 6.2 of
\Polchinski), and so don't contribute to
scattering from an ordinary D-brane.   However, such terms do seem to
contribute when considering D-branes with nontrivial worldvolume
fields, as is our interest.  Nevertheless, following standard practice,
we will continue to drop these terms by adopting this as our prescription
for defining the (naively divergent) $X^0$ path integral.}
\eqn\secbi{ \langle X'^\mu(z,\zb) X'^\nu(w,\wb) \rangle = \left\{ \eqalign{&
- {1 \over 2} \eta^{\mu\nu}\left[ \ln |z-w|^2 + \ln|1 - z \wb|^2 \right]
\quad N \cr & - {1 \over 2}
\eta^{\mu\nu}\left[ \ln |z-w|^2 - \ln|1 - z \wb|^2 \right] \quad
D.} \right.}
For $z$ in the bulk and $w$ on the boundary ($w= e^{it}$):
\eqn\secbj{  \langle X^{'\mu}(z,\zb) X^{'\nu}(t) \rangle = \left\{
\eqalign{ -
 \eta^{\mu\nu} \ln |z&-e^{it}|^2  \quad ~~~~N \cr & 0  ~~~~~~\quad\quad\quad D.} \right.}
For both points on the boundary:
\eqn\secbk{  \langle X^{'\mu}(t_1) X^{'\nu}(t_2) \rangle = \left\{
\eqalign{ -
 \eta^{\mu\nu} \ln |e^{it_1}&-e^{it_2}|^2  ~~~\quad N \cr & 0
~~~~~~\quad\quad\quad D.} \right.}
For self-contractions in the bulk:
\eqn\secbl{ \langle X^{'\mu}(z,\zb) X^{'\nu}(z,\zb) \rangle =
\left\{ \eqalign{&- {1 \over 2} \eta^{\mu\nu}
 \ln|1 - z \zb|^2 ~~\quad N \cr
&  ~~~~{1 \over 2} \eta^{\mu\nu}\ln|1 - z \zb|^2  ~~\quad D.} \right.}
For self-contractions on the boundary:
\eqn\secbm{ \langle X^{'\mu}(t) X^{'\nu}(t) \rangle = \left\{
\eqalign{&0  ~~\quad N \cr & 0   ~~\quad D.} \right.}

\subsec{Closed string tachyon scattering from static D-brane}

Using the above correlators, the  two point function of bulk
exponentials is \eqn\secbn{ \langle e^{i k_1\cdot  X'(z,\zb)}
e^{ik_2\cdot  X'(w,\wb)} \rangle =
 |z-w|^{ t} |1-z\wb|^{- s -  t} |1-z\zb|^{{ s\over 2}-2}
|1-w\wb|^{{  s \over 2}-2}. }
The S-matrix amplitude is given by fixing one vertex operator at the
origin, $z=0$, and integrating $w$ over the disk.  Up to an overall constant
and a momentum conserving delta function, the amplitude is
\eqn\secbp{ \int_0^1 \! dr ~ r~ r^{ t} (1-r^2)^{{ s \over
2}-2}
= { \Gamma\left( { t \over 2} +1 \right) \Gamma\left( { s \over 2}
-1 \right) \over \Gamma\left({ s \over 2}+ { t \over 2} \right)},}
which indeed exhibits poles in accord with \secbe.

In the remainder of this work we are interested in computing the
scattering amplitude from a decaying D-brane.  Physically, as a
function of $x^0$, we expect to find an amplitude which interpolates
between that of a D-brane and that of a collection of closed strings
(or tachyon matter) into which the D-brane decays. One signature of
this is that the open string poles should be absent at late times.
Besides this, it is difficult to exhibit any ``smoking gun'' signatures
of the presence/absence of the D-brane.  For instance, it is known that
in the high energy and fixed angle regime, both types of amplitudes
behave as $e^{-E^2 f(\theta)}$, indicating softness at short
distance.

\newsec{Scattering from the rolling tachyon}

\lref\HwangAN{ S.~Hwang, ``Cosets as gauge slices in SU(1,1)
strings,'' Phys.\ Lett.\ B {\bf 276}, 451 (1992)
[arXiv:hep-th/9110039];
J.~M.~Evans, M.~R.~Gaberdiel and M.~J.~Perry, ``The no-ghost
theorem for AdS(3) and the stringy exclusion principle,'' Nucl.\
Phys.\ B {\bf 535}, 152 (1998) [arXiv:hep-th/9806024].
}

Now we consider the tachyon profile $T(X^0)=\lambda e^{X^0}$ corresponding
to the boundary interaction
\eqn\nzmb{ e^{-S_{\rm bndy}} = e^{-\lambda e^{x^0} \int \! dt\, e^{X^{'0}} }}
where we have separated out the zero mode as in \ba.  Consider the
scattering of $\ell$ bulk tachyons.  The $X^0$ part of the
computation involves
\eqn\nzma{ \eqalign{A_{\ell}&= \int \! DX^0\, e^{-S}
\prod_{a=1}^\ell e^{i\omega_a X^0(z_a,\zb_a)} \cr &=\int\! dx^0\,
e^{i x^0 \sum_a \omega_a} \langle e^{-\lambda e^{x^0} \int \! dt\,
e^{X^{'0}} } \prod_{a=1}^\ell e^{i\omega_a
X^{'0}(z_a,\zb_a)}\rangle.}}
The full amplitude also contains terms involving the spacelike
$X^i$ that are the same as for a standard D-brane, as well as
integrals over the locations of the vertex operators $z_a$.    As
we will describe below the bulk vertex operators can be moved to
the boundary ($|z| = 1$), to compute amplitudes with arbitrary
numbers of open string tachyon vertex operators. We can always
choose a gauge in which our vertex operators (assuming they carry
nonzero energy) contain no timelike oscillators \HwangAN.
 Thus the interesting part of any correlation function
  (e.g., for closed string scattering from a brane with a perturbed initial state) involves
  interactions with the boundary tachyon that are summarized by \nzma.

\subsec{Perturbative approach and  matrix integral formulation}

One approach is to expand \nzma\ in powers of the boundary
interaction; {\it i.e.} as a power series in $\lambda$.  The magnitude of $\lambda$ can be
changed by shifting $x^0$, so
truncating the power series is  only sensible when
considering scattering processes which are dominated at early
times.  In general, we must keep and sum the
entire series. For the cases we consider, the sum will have a
finite radius of convergence.

Expanding, we obtain
\eqn\nzmd{A_{\ell}=\int dx^0 e^{i x^0 \sum_a \omega_a}
\sum_{N=0}^\infty {(-2\pi\lambda e^{x^0})^N \over N!} \int
\prod_{i=1}^N {dt_i \over 2\pi} \langle e^{X^{'0}(t_1)} \dots
e^{X^{'0}(t_N)}\prod_{a=1}^\ell e^{i\omega_a
X^{'0}(z_a,\zb_a)}\rangle.}
By separating out the $x^0$ integral in this way we are isolating the
contribution to the total scattering amplitude from the partially decayed
state of brane that is present at any particular time.
The late time contribution from $x^0 \to \infty$ should isolate
the effects of the tachyon matter final state to which the brane is
supposed to decay.

To calculate the fixed $x^0$ contributions we now need to evaluate
\eqn\nzme{  a_{\ell}^{(N)}=\int \prod_{i=1}^N {dt_i \over 2\pi}
\langle e^{X^{'0}(t_1)} \dots e^{X^{'0}(t_N)}\prod_{a=1}^\ell
e^{i\omega_a X^{'0}(z_a,\zb_a)}\rangle \ . }
The Wick contractions are straightforwardly evaluated using the
Green's functions in \secbi-\secbl, yielding

\eqn\zza{\eqalign{a_{\ell}^{(N)}&= \left[\prod_{a<b}
|z_a-z_b|^{-\omega_a \omega_b} \right]\left[\prod_{ab}
|1-z_a\zb_b|^{-\half \omega_a
 \omega_b}\right] \cr &\quad \times\int \prod_{i=1}^N {dt_i \over
2\pi}\left[\prod_{i<j}|e^{it_i}-e^{it_j}|^2\right]\left[
\prod_{ia} |1- z_ae^{-it_i}|^{2i \omega_a}\right]~.}}
An elegant way of rewriting \zza\ is in terms of matrix integrals
\LarsenWC\ConstableRC. In particular, the  measure $\prod_{i=1}^N
{dt_i \over 2\pi}\prod_{i<j}|e^{it_i}-e^{it_j}|^2$ is nothing but
the measure for $U(N)$ matrices expressed in the eigenvalue basis,
the product of exponentials being the Vandermonde determinant.
This leads to the identification
\eqn\zzb{a_{\ell}^{(N)}=N! \left[\prod_{a<b} |z_a-z_b|^{-\omega_a
\omega_b} \right]\left[\prod_{ab} |1-z_a\zb_b|^{-\half \omega_a
 \omega_b}\right]\int_{U(N)} \! dU \prod_a \left| \det (1-
 z_a U)\right|^{2i\omega_a}~, }
where is $U$ is a unitary $N\times N$ matrix, and we have
normalized the measure to $\int_{U(N)} \! dU =1$.

We can now write
\eqn\zzc{A_\ell =\left[\prod_{a<b} |z_a-z_b|^{-\omega_a \omega_b}
\right]\left[\prod_{ab} |1-z_a\zb_b|^{-\half \omega_a
 \omega_b}\right]\int \!dx^0 \,e^{i x^0 \sum_a \omega_a} F(z_1,\omega_1, \ldots ,z_\ell,\omega_\ell;\mu) }
where we have isolated the relevant summation by defining
\eqn\jab{F(z_1,\omega_1, \ldots ,z_\ell,\omega_\ell;\mu) =
\sum_{N=0}^\infty e^{-N\mu} \int_{U(N)} \! dU
 \,\prod_a
| \det (1 - z_a U ) |^{2i \omega_a}}
and
\eqn\jac{ \mu = -x^0 - \ln(-2\pi \lambda) \ . }
Here $\mu$ plays the role of a complex chemical potential, controlling the weights of the various $U(N)$ matrix integrals. As we
will discuss in Appendix D there is indeed a precise correspondence to the grand canonical ensemble of a statistical
mechanical system.  Note that the late time behavior corresponds to  ${\rm Re}(\mu) \rightarrow
-\infty$.

The essential problem is to compute the matrix integrals appearing
in \jab, perform the  sum over $N$, and evaluate the $x^0$
integral. Finally, to compute a string theory S-matrix element, we
must also integrate over the positions of the  vertex operators.

Having  obtained \zzc\ for general bulk tachyon amplitudes, it is
simple to include boundary tachyons.  Simply take any number of
vertex operators to the boundary, $z_a \rightarrow 1$, and remove
the corresponding factors of $|1- z_a\zb_a|^{-\half \omega_a^2}$,
since there are no boundary self-contractions according to \secbm.

\subsec{Integrated approach: performing the zero mode integral first}

An alternative approach is based on a strategy which was employed
with success in  bulk Liouville theory \GoulianQR. Instead of
leaving the $x^0$ integration until the end, we can perform it at
the outset. This approach requires analytic continuation in the
momenta and can miss non-analytic pieces that are known to be
present in the amplitude.   Nevertheless, useful information can
be obtained.

So let us return to \nzma\ and perform the $x^0$ integral using
\eqn\dab{\int_{-\infty}^\infty \! dx^0~ e^{-\alpha e^{x^0}} e^{i\omega x^0} =
\alpha^{-i\omega} \Gamma(i\omega)}
to obtain
\eqn\he{ A_\ell^{\rm int} = (2\pi \lambda)^{-i \sum_a \omega_a}
\Gamma(i\sum_a \omega_a)  \Big\langle \left[ \int \! {dt\over
2\pi}\, e^{X^{'0}(t)}\right]^{-i \sum_a \omega_a} \prod_{a=1}^\ell
e^{i\omega_a X^{'0}(z_a,\zb_a)}\Big\rangle \, .}
Using the identity $\Gamma(1-iz) \Gamma(iz) = -i\pi/\sinh{\pi z}$ we can write this as
\eqn\hea{\eqalign{
A_\ell^{\rm int}  &= -i\pi
{ (2\pi \lambda)^{-i \sum_a \omega_a}  \over \sinh(\pi \sum_a \omega_a)} \  B_\ell
\cr
B_\ell  &=
{1 \over \Gamma(1- i \sum_a \omega_a)}
 \Big\langle \left[ \int \! {dt\over
2\pi}\, e^{X^{'0}(t)}\right]^{-i \sum_a \omega_a} \prod_{a=1}^\ell
e^{i\omega_a X^{'0}(z_a,\zb_a)}\Big\rangle \, . } }
As we will see, the overall factor in $A_\ell^{\rm int}$ is of
exactly the form  expected for the closed string one-point
function \LambertZR.  The expression for $B_\ell$  contains
complex powers of the fields and therefore must be defined by
appropriate analytic continuation.   In particular, to apply
standard techniques, we need $-i \sum_a \omega_a$ to be a positive
integer.    We will now follow the procedure of defining $B_\ell$
by continuing the external momenta to imaginary integer values and
then ``continuing'' back. This procedure requires some care, since
in general the values of a  function at a discrete set of
arguments does not determine its behavior in the entire complex
plane even with an assumption of analyticity. However, we will
find situations in which the expression for $B_\ell$,
evaluated formally for imaginary integer momenta, is actually
defined on some open subset of the complex $\omega$ plane, thus
permitting reliable analytic continuation.

Proceeding in this fashion, we perform the continuation
\eqn\fa{ i\omega_a \rightarrow -n_{a} \ , }
to non-negative integers $n_a$, finding
\eqn\hf{
B_\ell = {1 \over N!}
\Big\langle \left[ \prod_{j=1}^N \int \! {dt_j \over 2\pi}
 e^{X^{'0}(t_j)} \right]
 \prod_{a=1}^\ell e^{-n_a X^{'0}(z_a,\zb_a)} \Big
\rangle}
with $N = \sum_a n_a$.    The correlator is the same as in \nzme\
after the substitution  \fa.  The result is therefore
 \eqn\zzd{
 B_\ell =
 \left[\prod_{a<b} |z_a-z_b|^{n_a n_b} \right]\left[\prod_{ab} |1-z_a\zb_b|^{\half n_a
 n_b}\right]\int_{U(N)} \! dU \prod_a \left| \det (1-
 z_a U)\right|^{-2 n_a}~.
 }
Comparing with \zzc\ an essential difference is that there is no
longer a sum over $N$ which needs to be evaluated, and the
exponents in \zzd\  are restricted to special integer values
(which we'll eventually have to continue to generic values).

At these imaginary integer momenta, and for sufficiently large
$N$, the matrix integral  \zzd\ is $N$ independent, and can be
efficiently evaluated group theoretically \ConstableRC\ or using
SU(2) current algebra \GutperleXF.
 As shown in  \refs{\ConstableRC},   for $N\geq \sum_a n_a$ (which applies
 to our case, since $N = \sum_a n_a)$,
\eqn\ekk{  \int \! dU \,\prod_a | \det (1 - z_a U ) |^{-2n_a} =
\prod_{ab}  |1 -z_a\zb_b|^{-n_a n_b}.}
This then yields
\eqn\el{
 B_\ell =
  \prod_{a<b} |z_a -z_b|^{n_a n_b}
\prod_{ab} |1-z_a\zb_b|^{-\half n_a n_b}.}

The final step is to ``undo" the analytic continuation, by
replacing  $n_a = -i\omega_a$ in \el, which, as we have stressed,
cannot be justified without additional information.  Nevertheless,
if we optimistically apply this procedure we are led to:
\eqn\ib{  A_\ell^{\rm int}= -i\pi {( 2\pi \lambda)^{-i \sum_a
\omega_a}\over \sinh( \pi \sum_a \omega_a ) } \prod_{a<b} |z_a
-z_b|^{-\omega_a \omega_b} \prod_{ab} |1-z_a\zb_b|^{\half \omega_a
\omega_b}~.}
In Sec. 4 we will see that naively using \ekk\  and then
analytically continuing to real momenta leads to unphysical
results for bulk-boundary amplitudes and we will describe a better
procedure for computing $B_\ell$.  In Sec. 5 we discuss different
procedures for computing bulk-bulk amplitudes.

\subsec{Comparison of one-point functions}

It is helpful to compare the results for the tachyon one-point
function ($\ell=1$) obtained in  the two approaches.\foot{This
computation also applies to an arbitrary closed string state,
since as pointed out in \LambertZR, we can always choose a gauge
such that there are no timelike oscillators.} By a conformal
transformation we can take the bulk vertex operator to be at the
origin of the disk, $z=0$.  In the perturbative approach we find
from \jab\ that $F =1/(1+2\pi \lambda e^{x^0} )$, and therefore
\eqn\zzg{ A_1 = \int\! dx^0 \,  {e^{i\omega x^0}  \over 1+2\pi
\lambda e^{x^0}}=-i\pi  {(2\pi \lambda)^{-i\omega}\over \sinh \pi
\omega}~.}
 Comparing with \ib\ we then find an agreement between the two approaches: $A_1 = A_1^{\rm int}$.

Note that in the perturbative approach the sum over $N$ is convergent only for $|2\pi \lambda e^{x^0}|<1$, but we define it in
general by analytic continuation.  The analytically continued sum vanishes exponentially for large $x^0$, which yields
 a convergent $x^0$ integral.

  The finite radius of convergence of the sum over $N$
is a feature of all the amplitudes we have examined, and makes it
challenging to extract  the late time behavior. In particular, to
find the late time behavior of the unintegrated amplitude we must
first find the {\it exact} early time behavior, so that we have an
exact formula to analytically continue. This in turn means that we
must perform an exact summation of the perturbation series. In the
calculation above this turned out to be easy, but the requisite
sums become  more challenging for higher point amplitudes.  In
light of this, it would be extremely useful to develop a method to
access the late time behavior directly without resorting to
analytic continuation.

Although we have so far focused on tachyon vertex operators, it is
instructive to consider a couple of other examples.
 Using the perturbative approach we can compute the one-point function of    $\p X^0 \overline{\p}X^0 e^{i\omega X^0}$ at the origin to
 be \LarsenWC\
\eqn\dja{\eqalign{ \langle e^{-S_{{\rm bndy}}}\p X^0
\overline{\p}X^0  e^{i\omega X^0}\rangle &= \int \! dx^0 \,
\left[\half {e^{i\omega x^0} \over 1+2\pi \lambda e^{x^0}}
-e^{i\omega x^0}\right]\cr & =-{i\pi \over 2}{(2\pi
\lambda)^{-i\omega}\over \sinh \pi \omega }-2\pi \delta(\omega)
.}}
The $\delta(\omega)$ terms  comes from the $N=0$ term in the sum,
which is special  since the bulk fields have nothing to contract
against.  On the other hand, if we repeat this calculation in the
approach where we first integrate over $x^0$ we will miss this
$\delta(\omega)$ term.  This is the because the latter approach is
based on analytic continuation in $\omega$ and can therefore miss
non-analytic pieces. The $\delta(\omega)$ term can be shown to be
responsible for energy conservation: it gives $T_{00}$ where
$T_{\mu\nu}$ is the energy-momentum tensor of the decaying brane.
The lesson is that the integrated approach can miss certain
physically relevant non-analytic contributions, but can still be
useful, provided we keep this lesson in mind.

Further subtleties are revealed when we study the one-point
functions of higher dimension operators corresponding to massive
string modes.  A finite number of terms in the sum over $N$ will
be ``special'', leading to integrands which diverge for late times
as $e^{px^0}$ for positive integer $p$ (this behavior was
observed in \OkudaYD, \OkuyamaJK.)  Such terms will also arise in
amplitudes involving more than one bulk tachyon, since such
operators will appear in the OPE.  This behavior clearly makes the
direct evaluation of $x^0$ integrals problematic.  To define these
integrals one should impose the constraints of conformal
invariance, as explained in \deBoerHD, \SenZM.  In particular, after
computing the tachyon one-point function, we can write the
remaining operators as Virasoro descendants, and then use the fact
that boundary state is conformally invariant to generate the
remaining amplitudes.

\newsec{Bulk-boundary two-point function: Perturbing the initial brane state}

A two-point amplitude in which one of the bulk operators is moved
to the boundary  describes an open-closed scattering amplitude.
The closed string could either be an in-state or out-state, but
because the D-brane is decaying and does not exist at late times,
the open string vertex operator can only be describing an
in-state.  Thus such amplitudes provide a systematic way of
exploring the effects of perturbations to the initial conditions
describing a decaying brane.  For example, we could construct a
localized lump on the brane at early times and ask how long in
time the effect of this deformation lasts.

In the previous sections, we have focused on closed string
tachyons. However,  the generalization to include an arbitrary
closed string state is relatively straight-forward if we choose a
gauge in which the on-shell closed string state with non-zero
energy has no time-like oscillators \HwangAN. In this gauge, the closed
string vertex operator has the form \eqn\csvo{V=e^{i\omega_c X^0}
V_{sp}~.} The spatial part of this operator $V_{sp}$ is constructed
from the 25 spatial fields and  is a Virasoro primary with
conformal dimension $\Delta=1 + {\omega_c^2 \over 4}$. The
computation of a bulk-boundary amplitude factorizes into a product
of the two point function in the time-like boundary Liouville
theory and in the free spatial CFT. The form of the closed string
vertex operator \csvo\ is useful, because in the time-like
boundary Liouville CFT the vertex operator has the same form as
the tachyon vertex operator. In the spatial direction, the
operator is non-trivial. We can write $V_{sp}$ as
\eqn\vsp{V_{sp}={\cal P}(\p \vec{X},\p^2 \vec{X} \cdots,\bar{\p}
\vec{X},\bar{\p}^2 \vec{X} \cdots)e^{i\vec{p}\cdot \vec{X}}~,} for a
polynomial  ${\cal P}$.
 In computing the bulk-boundary two point function with such an arbitrary
 closed string state, the contribution from the spatial part of the CFT  is given by
\eqn\spatial{\langle : {\cal P}(\p \vec{X},\p^2 \vec{X}
\cdots,\bar{\p} \vec{X},\bar{\p}^2 \vec{X}
\cdots)e^{i\vec{p}\cdot\vec{X}(0)}:~:
e^{i\vec{k}\vec{X}(t)}:\rangle~,} where $t$ parameterizes the
position of the open string vertex operator on the boundary. The
integration over the zero modes in the Neumann directions will
yield a factor of $\delta(\vec{p}_\parallel+\vec{k}_\parallel)$.
This $\delta$-function will be multiplied by terms from the
contractions of the bulk and boundary fields.  For no open string
vertex operator (or equivalently for $\vec{k}=0$) the result of
these contractions is just a phase \LambertZR. In particular, note
then that the absolute square of the result is independent of
$\omega_c$, even though $\omega_c$ enters into the definition of
$V_{sp}$ via $\Delta = 1 + {\omega_c^2 \over 4}$, a result which
is most easily seen from the boundary state.  In the case with the
open string vertex operator there will be additional contributions
from the various ways of contracting fields from the bulk and
boundary operators. This result will in general depend on the open
string momentum $\vec{k}$, and therefore on $\omega_o$.  However,
the squared result will again be independent of $\omega_c$ just as
before: $\omega_c$ dependence can potentially enter only in the
Dirichlet part of the computatation, but since the open string
vertex operator has no momentum components in the Dirichlet
directions, this part of the computation is identical to that
without the open string vertex operator \LambertZR.    So as far as the
spatial CFT is concerned, the effect of the open string vertex
operator is to contribute a $\omega_0$ and ${\cal P}$ dependent
factor, but not to alter the $\omega_c$ dependence.  Since our
primary interest is in the $\omega_c$ dependence, we will omit the
extra $\omega_0$ and ${\cal P}$ dependence (though these could be
straightforwardly computed), and so the formulas we will write
below are strictly valid only for ${\cal P}=1$, {\it i.e.} for the
closed string tachyon.


\lref\KrausCB{ P.~Kraus, A.~Ryzhov and M.~Shigemori, ``Strings in
noncompact spacetimes: Boundary terms and conserved  charges,''
Phys.\ Rev.\ D {\bf 66}, 106001 (2002) [arXiv:hep-th/0206080].
}

We now return to the computation of the bulk-boundary two point
function in the time-like  boundary Liouville theory. First
consider the perturbative approach.  To obtain a correlator with
one bulk and one boundary operator from \zzc, we set $\ell=2$, use
conformal invariance to place one operator at $z_1 = 0$ and
the other at $z_2=1$,\foot{Actually, this step is not entirely innocent.  
Separating out the zero mode $x^0$ breaks conformal
invariance, and so conformal invariance can only be restored after
performing the $x^0$ integration.  Since $x^0$ is noncompact,
there can be subtleties associated with boundary terms, as
discussed in \KrausCB.} and discard the divergent factor
$\left(1-z_2 \zb_2\right)^{-\half\omega_2^2}$ since it is removed
by boundary normal ordering. This gives
\eqn\nyac{ A_2(\omega_c,\omega_o) = \int \! dx^0 \, e^{i(\omega_c+\omega_o)x^0} F(0,\omega_c,1,\omega_o;\mu),}
with
\eqn\nybb{ F(0,\omega_c,1,\omega_o;\mu)= \sum_{N=0}^\infty e^{-N\mu}I_N(\omega_o) ~,}
and
\eqn\nybba{ I_N(\omega_o) = \int_{U(N)} \! dU \ |\det (1-U)|^{2i\omega_o} \ . }
Here $\mu$ is the ``chemical potential'' defined in \jac\ and
$\omega_{c,o}$ are the  energies of the closed and open string
vertex operators.

Now consider the integrated approach.   Following \hea , the integrated amplitude is
\eqn\zzia{A_2^{\rm int}(\omega_c,\omega_0) = -i\pi { (2\pi
\lambda)^{-i (\omega_c + \omega_o)}  \over \sinh(\pi (\omega_c +
\omega_o)}  \  B_2(\omega_c,\omega_o) \, . }
Here $B_2$ is defined by ``analytic continuation" from imaginary
momenta.  Setting $\omega_{c,o} = i n_{c,o}$ gives
\eqn\zzib{
B_2(in_c,in_o) = I_N(in_o)~,
}
with $N = n_c + n_o$.   Below we will evaluate $A_2^{\rm int}$
using both the integrated expression \zzia\ and the perturbative
expansion \nyac\ and show that they agree.  This bulk-boundary
amplitude captures the main features of more general amplitudes,
but in a cleaner context.    It allows one to vary the initial
state of the brane by creating an open string perturbation.  We
can therefore  explore a key physical question:  are the general
features of brane decay sensitive to the precise initial state
introduced by Sen.  For these reasons, we turn to a detailed study
of the bulk-boundary two point function.

\subsec{Use of the Selberg integral}

Our basic technical goal is to evaluate integrals of the form
\nybba.  This can be done with the help of a famous integral due
to Selberg. (For a pedagogical review, see \refs{\selbergs}.)
Following Sec.~2.1 of \refs{\snaith}
\eqn\ea{\eqalign{\!\!\!\!\int \! dU \, |\det(1- U)|^{-2\a} &=
{1\over N!}\int \!{dt_1 \over 2\pi}\cdots {dt_N \over 2\pi} \,
\prod_{i<j} |e^{it_i}-e^{it_j}|^2 \prod_i |1-e^{-it_i}
|^{-2\alpha} \cr &= {2^{N^2-2\a N} \over N! (2\pi)^N}
\int_{-\infty}^\infty \cdots \int_{-\infty}^\infty\! dx_1 \cdots
dx_N \, \prod_{i<j} |x_i-x_j|^2  \prod_{l=1}^N (1+x_l^2)^{-N+\a} \
. }}
The Selberg integral of interest to
us is \refs{\selbergs}
\eqn\eb{ \int_{-\infty}^\infty\! dx_1 \cdots dx_N \, \prod_{i<j}
|x_i-x_j|^2 \prod_{l=1}^N (1+x_l^2)^{-N+\a}= {(2\pi)^N N! \over
2^{N^2-2\a N}} \prod_{j=1}^N {\Gamma(j) \Gamma(j-2\a) \over
\left(\Gamma(j-\a)\right)^2}~.  }
We thus find that the integral relevant for the open-closed amplitude is
\eqn\ec{I_N(i\alpha) = \int \! dU \, |\det(1-U)|^{-2\a}
=\prod_{j=1}^N {\Gamma(j) \Gamma(j-2\a)  \over
\left(\Gamma(j-\a)\right)^2}~. }
Selberg's integral converges for real $\a$ when   $\a < \half$,
which is also the condition  for the Gamma functions on the right
side to have positive arguments. We will define the result for
arbitrary $\alpha$ by analytic continuation from the convergent
region.  In order to define the integrated amplitudes \zzia\ for
general real momenta we will have to also analytically continue
$N$.   The latter seems particularly problematic since the left
hand side of \ec\ is an integral over $U(N)$ matrices and the
right side contains a discrete product involving $N$.

Happily, progress can be made using the integral representation of the Gamma function, 
\eqn\intrep{ \log \Gamma(z) = \int_0^\infty {dt \over t} \left[ (z - 1) e^{-t} - {e^{-t} - e^{-zt} \over 1 - e^{-t}} \right]  , }
which is valid for $z > 0$.  (This domain of validity is the same
as that required for the convergence of the Selberg integral \eb,
namely $\alpha < 1/2$ for real $\alpha$.)    Applying this
identity gives the result:
\eqn\intrepa{
\log \left[ I_N(i\alpha) \right] = \log \left[B_2(i(N - \alpha),i\alpha) \right]
=
\int_0^\infty {dt \over t} (1 - e^{\alpha t})^2 { e^{-N t} - 1 \over 2(1 - \cosh{t})} ~.
}
Both $\alpha$ and $N$ can be analytically continued in this
expression.  We will use  this below to compute and compare the
perturbative \nybb\ and integrated \zzia\ 2-point scattering
amplitudes.    Although the integral expression is itself only
convergent  in some regions of the complex momentum plane, we will
see that it can be computed exactly in terms of special functions
in these regions which can then be continued to general values of
the momenta.

\subsec{Computing the Bulk-boundary amplitude}

\noindent {\bf $\bullet$ Integrated approach: }  In the approach
in which the zero mode is integrated at the outset   we wish to
evaluate \zzia.  We can do this by analytically continuing
\intrepa\ as $\alpha \to -i\omega_o$ and $N \to -i(\omega_c +
\omega_o)$.  This gives:
\eqn\zzj{\eqalign{A_2^{\rm int}(\omega_c,\omega_o)  =  &-i\pi {
(2\pi \lambda)^{-i(\omega_c + \omega_o)}\over \sinh \left( \pi
(\omega_c + \omega_o)\right)} e^{\int_0^\infty dt \, H(t,\omega_o)
\, (e^{i(\omega_c + \omega_o)t} -1) ~, } \cr H(t,\omega_o)\equiv &{
(1-e^{-i \omega_o t})^2  \over 2t(1- \cosh t) } ~.} }
 We will
evaluate  the integral in the exponent in closed form in terms of
special functions below.   But first let us compare \zzj\ to the
perturbative result \nybb.

\bigskip

\noindent{\bf $\bullet$ Perturbative calculation: }   In the
perturbative approach we must calculate the sum in \nybb.   In
terms of the function $H(t,\omega_o)$ defined above, this becomes
%
\eqn\pertcalca{
\eqalign{
F(0,\omega_c,1,\omega_0;\mu)
& = \sum_{N=0}^\infty e^{-N\mu}
e^{\int_0^\infty dt \, H(t,\omega_o) (e^{-tN} - 1)} \cr
&= \sum_{l,N=0}^\infty  {e^{-N \mu} \over l!}
\int_0^\infty \prod_{a=1}^l dt_a \, H(t_a,\omega_o) \, (e^{-N t_a} -1)
~.}
}
We can now expand out the product of $(e^{-t_a N} - 1)$ factors
in a series and explicitly carry out the sum on $N$.   This gives
\eqn\pertcalcb{
F(0,\omega_c,1,\omega_o;\mu) =
\sum_{l=0}^\infty
\sum_{m=0}^l
{(-1)^{l-m} \over l!} {l \choose m}
\int_0^\infty \prod_{a=1}^l dt_a \, H(t_a,\omega_o) \,
{1 \over
1 + 2\pi \lambda \, e^{x^0 - \sum_{b=1}^m t_b}
} ~,
}
where we have substituted back for the chemical potential $\mu$
in terms of the coupling $\lambda$ and the zero mode $x^0$.  The
full amplitude \nyac\ involves a Fourier transform of $x^0$ as in
\nyac.   We recognize the required transform from the expression
\zzg\ used for the computation of the closed string one-point
function and find
\eqn\pertcalcc{
\eqalign{
A_2(\omega_c,\omega_o) = -i\pi & {(2\pi\lambda)^{-i(\omega_c +
\omega_o)} \over \sinh(\pi (\omega_c + \omega_o))}
\sum_{l=0}^\infty \sum_{m=0}^l {(-1)^{l-m} \over l!} {l \choose m}
\cr & \times \int_0^\infty \prod_{a=1}^l dt_a \, H(t_a,\omega_o)
\, e^{i (\omega_c + \omega_o) \sum_{b=1}^m t_b} ~.} }
Carrying out the sums we arrive at the final result
\eqn\pertcalcd{
A_2(\omega_c,\omega_o) = -i\pi {(2\pi\lambda)^{-i(\omega_c +
\omega_o)} \over \sinh(\pi (\omega_c + \omega_o))}
e^{\int_0^\infty \! dt\,H(t,\omega_o) (e^{i (\omega_c +
\omega_o)t} -1)} ~.}
This precisely matches the integrated amplitude \zzj.\foot{Related results in the Liouville approach appear in \SchomerusVV.}  For future purposes, we define 
\eqn\Gdef{ G(\omega_c,\omega_o) \equiv {\int_0^\infty\! dt\,
H(t,\omega_o) (e^{i (\omega_c + \omega_o)t} -1)} } as the exponent
in the bulk-boundary amplitude.

The fact that the integrated and the perturbative approaches match
exactly greatly increases  our confidence in the methods developed
here.  The integrated approach is based on defining the
expressions in \hea\ by first expanding them for integer momenta,
evaluating the integrals and then interpreting the result as
definition of the amplitude for general momenta.  Such an
``analytic continuation" is inherently problematic -- without
further specification of asymptotic conditions there exist  many
analytic functions which coincide with the one at hand at integer
momenta.   By contrast, the perturbative approach does not suffer
from this failing and thus the agreement between \pertcalcd\ and
\zzj\ is very encouraging.  Nevertheless, both approaches  used
here involve analytic continuation in the momenta and run the risk
of missing non-analytic pieces.  In particular, the perturbative
approach assumed a well-defined Fourier transform and thus could
miss non-analytic contributions to the amplitude associated with
zero momentum processes.  Despite this we will be able to extract
useful insights into the structure of D-brane decay from
\pertcalcd.

\subsec{Comparison with other approaches}

As discussed in  the previous section, one might have proposed an
alternative approach to evaluating these amplitudes, by
analytically continuing $\omega_2$ to imaginary integer values
($-i \omega_2 = n \in Z^+$), since in that case \ConstableRC\
shows that
\eqn\eca{\int_{U(N)} \! dU \, |\det(1-zU)|^{-2n} =(1-z\zb)^{-n^2} ~.}
This expression  is valid for $n$ and $N$ nonnegative integers
with $n \leq N$.    Except for the trivial case $N=0$, the region
where this formula is valid is disjoint from the region $\a <
\half$ where the Selberg integral was valid prior to analytic
continuation in $\a$. To compare the two approaches, set $\a = n
\in Z^+$ in \ec\ and take $z=1$ in \eca\ to get the proposed
answer for the bulk-boundary amplitude. We find the following two
possible results
\eqn\ecb{ \int \! dU \, |\det(1-U)|^{-2n} = \left \{ \eqalign{&
\prod_{j=1}^N {\Gamma(j) \Gamma(j-2n)  \over
\left(\Gamma(j-n)\right)^2} = {\rm finite}~, \cr & (0)^{-n^2}
=\infty ~.} \right. }
The first expression is obtained from the Selberg formula used in
\ec\ while the second arises from \eca\ as derived in
\ConstableRC.    The Selberg result is finite for $N>2n$ and
 vanishes for $N \leq 2n$. By contrast,  the second formula diverges for
any $N$.  The consequence of this is that after continuation back to real momenta, naive use of the latter expression from \ConstableRC\ leads to a vanishing amplitude for all momenta, while the Selberg formula will lead to a finite result.

This situation can be given a useful worldsheet interpretation.
When a bulk vertex operator is taken towards the boundary to
define a bulk-boundary amplitude, it will collide with operators
from the boundary interaction. Therefore, we need to add
counterterms to operators which are taken to the boundary in order
to dress them appropriately in the presence of the interaction.
The second formula  in \ecb\ corresponds, however, to a bare
vertex operator, and so diverges. On the other hand, the analytic
continuation used to define the Selberg integral is  a convenient
way of regulating and renormalizing the operator, and so gives a
finite result.  Of the two results displayed in
\ecb, the finite Selberg result is the physically relevant one.

The summation over $N$ that  was carried out in the perturbative
approach, and which led to \pertcalcd\ for the final amplitude,
became possible because the Fourier transform in the definition of
$A_2$ \nyac\ simplified the expressions.  However, it can also be
of interest to do the sum on $N$ in \nybb\ {\it without}
integrating over the zero mode $x^0$.  As we discussed earlier,
the contributions to the scattering amplitude from late times
(large $x^0$) could shed  light on the nature of the ``tachyon
matter'' state to which the brane is supposed to decay.  Doing the
perturbative sum in \nybb\ exactly for fixed small $x^0$ and then
continuing to large $x^0$ turns out to be very difficult for
arbitrary momenta. However, it is possible to make progress for
special imaginary integer momenta. The relevant techniques and
results are presented in Appendix A.

\subsec{Initial state perturbations and the decay of the brane}

Our result \zzj\ (or \pertcalcd) for closed string emission  from
a decaying brane perturbed by an additional open string tachyon
allows us to explore how the emission of closed strings is
modified when the initial conditions for the brane are changed.
For example, we might make a tachyon lump on the brane by
superposing open string vertex operators and ask how this affects
the decay.

Important questions to answer are whether all the energy  in the
brane decays into closed strings, and what is the distribution of
these strings.   Sen has argued that brane decay leads to a
pressureless state of ``tachyon matter''\SenIN. A computation of
closed string emission from decaying branes by \LambertZR\
(summarized in eq. \zzg\ in Sec.~3) showed that at tree level the
emission of closed strings is exponentially suppressed in their
energy.  For low dimension branes this nevertheless implies a
divergent total emission, because of the Hagedorn growth in the
density of states. This divergence was interpreted as indicating
that all the energy of the decaying brane would be converted into
closed strings, and that very heavy strings would dominate the
decay products.   Such heavy strings would have the stress tensor
of pressureless dust, thus suggesting that they constitute the
mysterious tachyon matter. It was also argued that near the
endpoint of the decay, including back-reaction would ensure that
energy conservation was satisfied so that the emission of closed
strings shuts off after all the energy in the original brane has
been converted.  This was confirmed in the $c=1$ matrix model,
where it was also found that it is crucial to take into account
the quantum mechanical nature of the decaying brane \KlebanovKM.
By contrast higher dimensional branes had a finite total emission
of energy. The conventional wisdom states that in this case, small
perturbations can also lead to decay into higher co-dimension
branes \LarsenWC\ which would account for the ``missing energy''
in the decay into closed strings.

Our results can enable  a systematic exploration of how changing
the brane initial conditions affects the decay products.  To
initiate this study we will compute a closed form expression for
the closed string emission amplitude \zzj\ in terms of special
functions and then extract the asymptotics of the decay for large
closed string energies.

Before plunging into the calculation, let us note that there are
two possibilities for what one might mean by ``changing
the brane initial conditions''.   One interpretation involves
describing the brane by a boundary state; then to change the
boundary state we can act with an exponentiated open string vertex
operator (at least to first order in the operator; beyond that
there will be corrections).  This corresponds to changing the
classical open string tachyon profile of the decaying brane. To
first order in the perturbation, the total emission rate into
closed strings will then be the sum of the original rate plus a
correction due to the perturbation.    Here we will focus on an
alternative interpretation, in which we add a single quantum open
string excitation on top of the classical background. The
open-closed amplitude then corresponds to the amplitude for the
incoming open string to disappear and for a closed string to be
created.  

To compute the  bulk-boundary amplitude \zzj\ in closed form we must evaluate an integral of the
form
\eqn\Jdef{ G(\omega_c,\omega_o) = \int_0^\infty  {dt \over 2t} \,
(1 - e^{-i\omega_o  t})^2 \,  {e^{-\beta t} - 1 \over  (1 - \cosh
t)} ~~~~;~~~~ \beta = -i(\omega_c + \omega_o)~.
}
%
Notice first that this integral vanishes if  $\omega_o = 0$, i.e.,
in the absence of any open string perturbation.  In that case,
\eqn\openzero{
A_2(\omega_c,0) =
-i\pi {(2\pi\lambda)^{-i\omega_c} \over \sinh(\pi\omega_c)} ~,
}
 reproducing the known result in \LambertZR.  So the full result
has the form of a modulation of the emission amplitude without an
open string perturbation by a factor of $e^G$.   It is also easy
to argue that the real part of $G$ will be negative so that the
modulating factor suppresses the amplitude.

The  challenge in evaluating the integral \Jdef\ occurs because as
$t \to 0$, cancellations between the terms in the numerator are
needed to cancel the overall power of $t^3$ in the denominator in
this limit, leading to a finite integral.  With this in mind, to
evaluate $G$ for general momenta,  observe that $G$ also vanishes
if $\beta = i(\omega_c + \omega_ o) = 0$.  Because of this, we can
write
\eqn\Kdefa{
G(\omega_c,\omega_o) =
 \int_0^\beta  d\beta \, {\partial G\over \partial \beta}\Big{|}_{\omega_o}
\equiv   - \int_0^\beta d\beta \, \int_0^\infty {dt \over 2} \, (1
- e^{-i\omega_o  t})^2 \,  {e^{-\beta t}  \over  (1 - \cosh t)} ~.
}
After  expanding $(1 - e^{i \omega_o t})^2$, each term in
$\partial G / \partial \beta$ can be evaluated with the help of
the integral formula
\eqn\bateman{\int_0^\infty \Bigl(1-\cosh t \Bigr)^\nu e^{-\gamma t} dt =e^{i\pi \nu} 2^{-\nu}
 {\Gamma(\gamma -\nu) \Gamma(2\nu + 1) \over \Gamma(\gamma + \nu +1)}~;
 ~~~~~~ {\rm Re} \nu > -{1 \over 2}~,}
which is listed in \Bateman\ (vol. 1, p. 163).   This  expression
diverges as $\nu \to -1$, but the linear combination \eqn\Kdefc{
-e^{-i\pi\nu} 2^{-\nu-1} \Gamma(2\nu+1) \Big\{ {\Gamma(\gamma -
\nu) \over \Gamma(\gamma + \nu +1 )} - 2 {\Gamma(\gamma - \nu + i
\omega_0) \over \Gamma(\gamma + \nu +1 + i\omega_o)}
+{\Gamma(\gamma - \nu + 2i\omega_o) \over \Gamma(\gamma + \nu +1 +
2 i \omega_o )} \Big\} }
needed here is well-defined as $\nu \to -1$.

The limit $\nu \to -1$ gives $\partial G/\partial \beta$ in  terms
of derivatives of the Gamma function.   The integral with respect
to $\beta$ that remains can be evaluated by again using the
integral representation \intrepa.  The result can be written in
terms of the Hurwitz $\zeta$ function
\eqn\Hurwdef{ \zeta(s,z) = { 1 \over \Gamma(s)} \int_0^\infty
{t^{s-1} e^{-zt}  \over 1-e^{-t}}= \sum_{n=0}^\infty { 1\over
(n+z)^s} ~,}
and $\zeta^{m,n}(s,z)\equiv {\partial^{m+n} \zeta(s,z) \over
\partial s^m \partial z^n}$.  In order to summarize the results,
we introduce the notation
\eqn\Hdef{
\CH[F(x),a,b)] \equiv F(b+a) - 2 F(b) + F(b-a)~.
}
Then, after using various $\zeta$ function identities, we find the result:
\eqn\Gfinal{\eqalign{ G(\omega_c,\omega_o) & =
\CH[\zeta^{(1,0)}(-1,x),-i\omega_o,i\omega_o] -
\CH[\zeta^{(1,0)}(-1,x),-i\omega_o,-i\omega_c]  \cr
    & +    \CH[x(\zeta(0,x) - \ln\Gamma(x)),-i\omega_o,i\omega_o]  - \CH[x(\zeta(0,x) -\ln\Gamma(x)),-i\omega_o,-i\omega_c]~.
}
}
Putting this expression back into \pertcalcd\ gives
the general  bulk-boundary amplitude in closed form in terms of
zeta functions.   This result can now used to extract many aspects
of the physics of decaying branes.  Here we will explore one
question -- how does including an open string perturbation change
the asymptotics of brane decay into closed strings?

To do this we must find the asymptotics of $G(\omega_c,\omega_o)$
for $\omega_c \gg \omega_o$.    We will use the expressions
\eqn\zetaasymp{\eqalign{
\zeta(0,x) & = {1 \over 2} - x~, \cr
\zeta^{(1,0)}(-1,a) & =
{B_2(a) \ln{a} \over 2} - {B_2(a) + a \over 4} + {B_2 \over 4}
 - \sum_{k=3}^\infty {a^{-k+2} B_k \over (k-2) (k-1) k}
 ~.}}
Here $B_2(a) = a^2 -a + 1/6$ and $B_2 = 1/6$ are  the second
Bernoulli polynomial and number respectively.  The second formula
was derived for $\Re(a) \to \infty$ \Mathworld,
but it will apply by analytic continuation for large $|a|$ in
general.   Applying these formulae we find that for $\omega_c \gg
\omega_o$, the dominant term is $G(\omega_c,\omega_o) \approx
-\omega_o^2 \ln \omega_c$.    If $\omega_o$ is itself large the
formula is
%
\eqn\Gdom{
G(\omega_c,\omega_o)
\approx - \omega_o^2 \ln\Big( {\omega_c \over \omega_o} \Big) ~.
}
For small $\omega_o$ and $\omega_c \gg \omega_o$ the denominator
in the log is modified, but the $-\omega_o^2 \ln \omega_c$
behaviour always holds. The next to leading terms are independent
of $\omega_c$, but dependent  on $\omega_o$ and include a phase.
Beyond that terms are suppressed by powers of $\omega_c$.  We have checked these
asymptotics numerically.

To summarize, the amplitude for emitting high energy closed
strings, when the brane initial state has been perturbed by a
reasonably energetic open string which is subsequently absorbed,
is
\eqn\approxampla{
A_2(\omega_c,\omega_o) \approx
-i\pi {(2\pi\lambda)^{-i(\omega_c + \omega_o)} \over \sinh(\pi (\omega_c + \omega_o))}
\Big({\omega_o \over \omega_c}\Big)^{\omega_o^2}  \cdots
}
where the ellipses represent terms that are either constant or fall off as $\omega_c$ increases.

For the unperturbed initial state of the D-brane (the spatially
homogeneous decay), it was found in \LambertZR\ that the total
energy of closed strings emitted was divergent for a Dp brane with
$ p \leq 2$ (but see also  \UVfinite). We now examine how this statement is modified when
the initial state is perturbed by addition of the boundary tachyon
vertex operator.  To compute the expectation value of total
emitted energy we add up the squares of the individual emission
amplitudes,
%
\eqn\eperv{{E
\over V_p}\sim \sum_s {1 \over 2}|A^{(s)}_2(\omega_c,\omega_o)|^2 ~.}
For large $\omega_c$ \eqn\largee{|A_2(\omega_c,\omega_0)| \sim
e^{-2\pi \omega_c} {\omega_c}^{-\omega_0^2 ~}~.}
 This differs from the
unperturbed case in the extra factor of $\omega_c^{-\omega_o^2}$.
A similar calculation to the one done in \LambertZR\ yields
\eqn\epervb{{E \over V_p} \sim{1 \over (2\pi)^p} \int
\!d\omega_c\, \omega_c^{-{p \over 2}-\omega_o^2}~.} Notice that the
extra factor of $\omega_c^{-\omega_o^2}$ is a suppression if
$\omega_o$ is real but an enhancement for imaginary $\omega_o$.
Without the initial state perturbation, the average energy per
unit volume diverges for the D0, D1, and D2 branes. For the D0
brane, the on shell condition fixes the  initial state
perturbation with the open-string tachyon vertex operator to be
$e^{X^0}$, i.e. $\omega_o=i$ and the perturbation enhances the
closed string emission leading to a divergent result. For the D1
brane, the result depends on the details of the perturbation. If
the spatial momentum is large enough, such that $\omega_0 > 1$,
the closed string emission will be finite.   For higher dimensional branes, the total emission will be finite for any perturbation with real $\omega_0$.  If the spatial
momentum on any brane is sufficiently low, the on-shell condition
will cause $\omega_o$ to be imaginary, and there will be an
enhancement of closed string emission.  For $p \leq 3$ the energy contained in closed string emission will diverge in this case.

 \subsec{Comments}
 
For the D0 brane, the divergent closed string emission is expected to be
an artifact of perturbation theory, which will be regulated
by back reaction and quantum effects. It strongly suggests that
the initial D-brane decays completely into closed strings and that
tachyon matter, the end product of the decay in classical open
string theory, is made up of very heavy closed strings \LambertZR.  A similar interpretation is given to the divergent emission of closed strings from the unperturbed D1 and D2 branes. 
For higher dimensional branes ($p \geq 3$) the conventional wisdom has been
that the decay process is more elaborate since it
proceeds inhomogeneously (\LarsenWC, \SenVV, \MukhopadhyayEN).     Notice, however, that in a quantum mechanical treatment of a decaying brane of {\it any} dimension, there will always be a component of the brane wavefunction along operators with imaginary $\omega_o$.   This is because such an open string operator will be proportional to $e^{|\omega_o| X^0}$ and so be very small at early times, in effect corresponding to an infinitesimal  perturbation of the initial state.  We have shown that such perturbations enhance closed string emission for all Dp-branes, even when the decay is homogeneous, and would lead to divergent emission (without treating backreaction and wavefunction effects) for $p \leq 3$. In fact, we will get a divergent emission for any $p$ if we are willing to consider non-normalizable perturbations with imaginary spatial momenta.   This enhancement of closed string emission is appealing.  It suggests that a full treatment of decaying branes of any dimension will show all the energy of the brane entering into a state of heavy closed strings.

In our study of the open-closed amplitude we have discussed the open string state as an in-state, or a perturbation of the initial condition, while the closed string is emitted as an out-state.  By complex conjugating we could have naively written down an amplitude that would apparently describe the emission of an open string.  Likewise, at the perturbative level, we could have computed amplitudes containing additional strings in the final state including, apparently, open strings.   This seems puzzling given that the brane has decayed at late times and thus open strings should not exist.  It is possible that such amplitudes are simply inconsistent at a non-perturbative level.  However, it seems more reasonable that we should interpret such amplitudes with ``open strings'' in the final state as follows.  In any time dependent background we construct vertex operators at early and late times (in- and out- operators) to describe simple perturbations around the respective vacua.   However, we are always free to use the complete basis of  in-operators to describe outgoing states also -- these operators will simply describe very complicated out-states.   In our context, both open and closed strings can exist at early times, but only closed strings exist at late times.   An out-open-string operator should thus represent some very complicated correlated state of closed strings that emerges from the open string via the decay of the brane.    This interpretation leads to the important question of whether, in order to compute the full decay of a brane, we must calculate amplitudes with all possible out vertex operators, or whether computing the emission of closed strings as in \LambertZR, or as done above, suffices.    In other words, do we include all decay channels that are perturbatively apparent including the ones with final-state ``open strings'', or is this ``double counting''?  The answer to this question goes to the heart of the open-closed duality for decaying branes that has been conjectured by Sen (\SenXS, \SenIV).   Regardless, the amplitudes computed above are essential components of the full answer and are the complete result for the ``exclusive'' amplitude describing absorption of the initial open string and emission of the final closed string.

\newsec{The bulk-bulk amplitude}

Now we turn to a computation of the bulk-bulk amplitude which
describes the scattering of a closed string from the decaying
brane, or the emission/absorption of a correlated pair of closed
strings during the decay.  To obtain the relevant correlator from
\zzc\ and \jab\ we use the conformal symmetry  (but recall
footnote 10) to locate one vertex operator at $z=0$ and the other
at $z = r$, $0\leq r \leq 1$ on the unit disk. This gives
\eqn\ampa{ A_2(\omega_1,\omega_2,r) = |r|^{-\omega_1\omega_2} \,
(1 - |r|^2)^{-\omega_2^2/2} \, \int \!dx^0 e^{ix^0 (\omega_1 +
\omega_2)} \, F(0,\omega_1,r,\omega_2;\mu) ~,}
with $\mu = -x^0 -\ln(-2\pi\lambda)$ and
\eqn\ampb{\eqalign{
F(0,\omega_1,r,\omega_2;\mu) & =
\sum_{N=0}^\infty e^{-N \mu} J_N(\omega_2,r)~, \cr
J_N(\omega_2,r) & = \int_{U(N)} dU \, |\det(1 - rU)|^{2i\omega_2} \, .
}
}
To compute the full amplitude we should integrate over $r$ as
$\int_0^1 \! dr \,r$.

\subsec{The Toeplitz determinant}

%
The key steps in computing the amplitude are to evaluate the terms
$J_N$ in the infinite series \ampb\ and carry out the sum. Each
$J_N$ is an expectation value of a periodic function with respect
to the circular unitary ensemble (Appendix B):
\eqn\amplc{\eqalign{J_N(\omega_2,r) & = {1\over N!}
\int^{\pi}_{-\pi}\prod_a {dt_a \over 2\pi }
 \prod^n_{a=1} |1-re^{-it_a}|^{2i\omega_2}
 \prod_{a<b} |e^{it_a}-e^{it_b}|^2  \cr
 & \equiv \Ebf_N \Bigl\{ \prod^N_{i=1}f(t_i) \Bigr\}  \ . }}
By Heine's identity \Heine, this expectation value is equal to the
Toeplitz determinant of the Fourier coefficients of $f$,
\eqn\ampld{\Ebf_N \Bigl\{ \prod^n_{i=1}f(t_i) \Bigr\} =
D_N[f]
 \equiv    \det (\hat{f}_{j-k})_{1\leq j,k\leq N}
~. }
%
By this notation  we mean the determinant of a matrix in which the
entry in the $j^{{\rm th}}$ row and  $k^{{\rm th}}$ column is the
$(j-k)^{{\rm th}}$ Fourier coefficient of $f$.   For reference,
the proof of this identity is included in the Appendix B.  In our
case the Fourier coefficients are (denoting $\alpha =-i\omega_2$)
\eqn\amplf{\hat{f}_k= \hat{f}_{|k|} =
\int {dt \over 2 \pi} |1-re^{-it}|^{-2\alpha}
e^{-ikt} = {r^{|k|} \over |k| B(\alpha ,|k|)}
F(\alpha , |k|+\alpha, |k|+1;r^2) \ .}
Here $B$ is a beta function and $F$ is a hypergeometric function.
This integral representation of the hypergeometric function is
valid for $\alpha \neq 0, -1, -2\cdots$ and $|r| < 1$
\GradshteynRyzhik. As we will see below, the boundary limit $r \to
1$ of the right hand side will be well-defined anyway.  The
restriction on $\alpha$ can also be written as $\omega_2 \neq 0,
-i, -2i, -3i \cdots$.

\medskip

\noindent {\bf $\bullet$  Special imaginary momenta: }   When
$\omega_2 = i, \, 2i, \, 3i \cdots$ (i.e., positive imaginary
integer momenta), or equivalently $\alpha = 1, 2, 3 \cdots$ with
$\alpha \leq N$ we know that the integral $\ampb$ could have been
done by the group theoretic methods of \ConstableRC\ with the
result that
\eqn\Jspecial{
J_N(\omega_2,r) = (1 - r^2)^{\omega_2^2} ~~~;~~~~
\omega_2 = in,~n \in Z^+ ~~~;~~~ n \leq N
\, .
}
Since these momenta are within the range of validity of  the
integral representation \amplf, the determinant $D_N[f]$ in
\ampld\ must reproduce \Jspecial\ at positive imaginary integer
$\omega_2$.  It is easy to verify that this is indeed the case, at
least for small $n$ and $N$ but we have not constructed a general
proof.

\medskip

\noindent {\bf $\bullet$ Boundary limit gives Selberg: }  As $r^2
\rightarrow 1$, the integrals and coefficients \ampld \ reduce to
\ec\ of the previous bulk-boundary calculation.
In the limit, the determinants \ampld \ reduce to
\eqn\amplea{D_N[f]_{r=1} = \Bigl( {\Gamma (1-2\alpha ) \over
(\Gamma (1-\alpha )) \Gamma(\alpha)} \Bigr)^N \det \bigl\{
{\Gamma (\alpha + |i-j| )
\over \Gamma (1+|i-j|-\alpha )} \bigr\}_{i,j=1,\ldots ,N} \ .}
Hence this determinant must be equal to \ec:
\eqn\ampleb{J_N(\omega_2,1) = D_N[f]_{r=1} = \prod^N_{j=1} {\Gamma (j) \Gamma (j-2\alpha)
\over (\Gamma (j-\alpha ))^2 } \ . }
We  have checked the identity explicitly with Maple for $N=1,\dots
,8$, but have not constructed a general proof.

These  two special limits appear initially to be in contradiction
with each other since, for $\omega_2 = in$, \Jspecial\ diverges as
$r \to 1$, while \ampleb\ is finite.  (Also see the discussion
around \ecb.)  However, what we really seeing here is that the
order of limits can matter when one approaches the boundary at $r
\to 1$ at special momenta.   Our Toeplitz determinant \ampld\
provides a definition of the two point amplitude for general
momenta and gives a well-defined analytic continuation to the
bulk-boundary amplitudes (with $r=1$) that we derived in the
previous section.   In addition the determinant reproduces the
special form \Jspecial\ which was previously obtained at imaginary
integer moment \ConstableRC. Therefore, it appears that naive
analytic continuation of \Jspecial\ to general $\omega_2$, which
is the technique used in the past, is invalid, as one might have
expected.

Below  we will see that the naive continuation of \Jspecial\ from
imaginary integers to general $\omega_2$ amounts to a large N
approximation.

\subsec{Large N approximation}
In general, dealing with the Toeplitz determinants
is very complicated. We will consider two methods.
For large $N$, approximate results can be found by
using Szeg\"o's limit theorem. This estimate can be improved
by using an  identity due to Borodin and Okounkov
\BorodinOkounkov \ (see also \BasorWidom ).
The latter authors considered Toeplitz determinants
$D_n[f]$ of periodic functions and related them to Fredholm determinants
of certain kernels. In particular,  they considered functions
of type
\eqn\amplm{f=(1+\xi_1 \eta)^{z}(1+\xi_2 \eta^{-1})^{z'} \ , }
where $\xi_1,\xi_2,z,z'$ are complex parameters and $\eta$ is a
coordinate on a complex plane.  If we set $\eta = -e^{-it}$,
$\xi_1=\xi_2=r$, and $z=z'=-\alpha$, we recover our particular
case. Here $\alpha$ can be an arbitrary complex number.  Following
\BorodinOkounkov, there is an identity
\eqn\ampln{D_N[f]=Z~\det (1-K_N) ~ .}
The prefactor $Z$ involves the Fourier coefficients of $\ln f$:
\eqn\amplo{Z = \exp \bigl( \sum^{\infty}_{k=1} k
\widehat{(\ln f)}_k\widehat{(\ln f)}_{-k}
\bigr) \ . }
This is the large $N$ approximation for the
determinant given by Szeg\"o's limit theorem,
$\lim_{N\rightarrow \infty} D_N[f]=Z$, whereas the
factor $\det (1-K_N)$ contains all the corrections to the approximation.
To evaluate $Z$  we use the relation  \GradshteynRyzhik\
(assuming $0\leq r^2\leq 1$)
\eqn\ampli{
\widehat{(\ln f)}_k
= -{\alpha \over 2\pi}  \int^{\pi}_{-\pi}
\ln (1-2r\cos(t) +r^2) e^{-ikt}dt
= -\alpha {r^{|k|}\over |k| } \, ,
}
which gives
\eqn\amplj{ D_N[f] \approx Z  =  \exp \{ \alpha^2 \sum^{\infty}_{k=1}
k~r^{2k}k^{-2}\}  = (1-r^2)^{-\alpha^2} \ . }
This  large $N$ approximation reproduces the  naive
analytic continuation $n^2 \rightarrow \alpha^2$ of  \ekk\ which
results from calculations at imaginary integer momenta
\ConstableRC.   In fact, as we showed in \Jspecial, when $\alpha
= n,~n\in Z^+,~n\leq N$, the result \amplj\ is {\it exact}, or
equivalently, the Borodin-Okounkov determinant $\det(1-K_N)$
equals 1.   There appears to be a remarkable localization
phenomenon in the amplitude \ampb\ such that for special momenta
the result localizes to the large $N$ result $Z$ \amplo.

 Notice that the right hand side of \amplj\ is independent of the dimension $N$ of the matrix.
 It is   dangerous to use this approximate large $N$ result in the bulk-bulk amplitude
  \ampb because of the infinite sum, but if we do so regardless, we get
  (restoring $\alpha =-i\omega_2$)
\eqn\amplk{
F(0,\omega_1,r,\omega_2 ;\mu )  = \sum^{\infty}_{N=0}
e^{-N\mu}
(1-r^2)^{\omega_2^2}
= {(1-r^2)^{\omega_2^2} \over 1 + 2\pi \lambda e^{x^0} } \ .
}
Putting this back in the complete amplitude \ampa\ and integrating over $x^0$ gives
\eqn\compampa{
A_2(\omega_1,\omega_2,r) =
-i\pi { (2\pi\lambda)^{-i(\omega_1 + \omega_2)} \over \sinh(\pi(\omega_1 + \omega_2))}
\,
 |r|^{-\omega_1\omega_2}  (1 - |r|^2)^{\omega_2^2/2} ~.
}
This gives the same result as in \ib, computed by doing the zero
mode integral  first and using naive continuation from large
imaginary integer momenta.   The full amplitude is computed by
including the spatial part of the amplitude and integrating over
the position of the vertex operator.  To do this recall that the
spatial part of the conformal field is unchanged by the boundary
perturbation, so we can include the contributions for a standard
D-brane in \secbn, and write
\eqn\compampb{
{\bf A}_2 = \int_0^1 dr \, r \, A_2(\omega_1,\omega_2,r) =
-i\pi { (2\pi\lambda)^{-i(\omega_1 + \omega_2)} \over \sinh(\pi(\omega_1 + \omega_2))}
\int_0^1 dr \, r \, |r|^{t} (1 - |r|^2)^{ s/2 - 2 + \omega_2^2}~,
}
where we used the Mandelstam variables \secbc\ and the closed
string tachyon on-shell  condition (\secbd\ with $N=0$).
Finally, integrating over $r$ gives the result
\eqn\ampll{
{\bf A}_2 =
-i\pi { (2\pi\lambda)^{-i(\omega_1 + \omega_2)} \over \sinh(\pi(\omega_1 + \omega_2))}
\,
{\Gamma ( (t/2)+1) \Gamma (\vec{k}^2_{||}-1) \over \Gamma
((t/2)+\vec{k}^2_{||} )}~,
}
where $\vec{k}_{||}$ is the spacelike part of the momentum in the directions parallel to the brane.
%
The factor $\Gamma(t/2 + 1)$ in this expression contains the
closed string poles \secbd.   The remaining poles coming from the
$\Gamma(\vec{k}_{||}^2 - 1)$ are more exotic, since they occur at
special spacelike momenta.  If momentum here was not simply the
spacelike piece, we would be recovering  open string poles.

Above we pointed out that the boundary limit $r \to 1$ of the full
amplitude \ampleb\ gives the result of the Selberg integral that
we used to compute the bulk-boundary amplitudes in Sec.~4.   By
contrast, the large $N$ approximation described here, since it
coincides with the naive continuation from imaginary integer
momenta, gives a singular boundary limit.  (See \ecb\ and the
discussion around it.)    In Sec.~4.3 we argued that the singular
behavior of the naive analytic continuation could be related to a
need to regulate and renormalize bulk operators as they approach
the boundary and collide with the boundary perturbation.  Here we
have seen that a large $N$ approximation of the bulk result leads
to a singular boundary limit, while the full expression leads to
the finite quantity that was used in Sec.~4 to study the effect
of initial state perturbations on the decay of the brane. It may
be useful to think of the large $N$ approximation \amplj\ as a
saddlepoint contribution, with the Borodin-Okounkov determinant in
\ampln\ containing corrections due to fluctuations in the
interactions between the vertex operators and the boundary
perturbations. Perhaps it is possible to understand the
determinant in terms of a renormalization or dressing of string
vertex operators by the boundary interaction.

Our result demonstrates that the  usual technique for computing
decaying brane amplitudes by analytically continuing from
imaginary integer momenta is unreliable.   The naive continuation
only isolates a large N  contribution to the full answer. This
happens because of the remarkable localization of the full
integral to the large N contribution at imaginary integer momenta.
Our amplitudes are directly defined at real momenta and differ in
important ways from the naive continuation. We have already
discussed the consequences of the extra contributions for
open-closed amplitudes.  We now turn to a brief examination of the
extra factor, the Borodin-Okounkov determinant, in a general
amplitude.

\subsec{The Borodin-Okounkov determinant}

The determinant in the full amplitude $D_N[f] = Z \, \det(1 -
K_N)$ can be expressed  (\BorodinOkounkov\ and \BasorWidom) as  a
Fredholm determinant of an operator $K_N$, acting on the $l_2$
normed infinite vector space $l_2(N,N+1,\ldots)$, defined by
\eqn\amplp{\det(1-K)=1+\sum^{\infty}_{m=1}(-1)^m
\sum^{\infty}_{N\leq l_1<l_2<\cdots <l_m} \det[K(l_i,l_j)]^{m}_{i,j=1} \ .}
This equation is essentially a discrete version of the
Fredholm determinant of a kernel familiar from the theory of
integral equations.
For the particular case of \amplm\ (with the parameters
replaced by ours), the kernel $K$ takes the form (correcting a minor
error in \BorodinOkounkov)
\eqn\amplq{\eqalign{K(i,j) &= {(-\alpha )_{i+1}(-\alpha )_{j+1} \over i!j!}
r^{i+j+2}(1-r^2)^{-2\alpha -1} \cr
& \times {1 \over i-j} \bigl( F(\alpha ,-\alpha ,i+1,{r^2 \over r^2-1})
{F(-\alpha -1 ,1+\alpha ,j+2;{r^2 \over r^2 -1}) \over j+1} \cr
& - {F(-1-\alpha ,1+\alpha ,i+2; {r^2 \over r^2 -1 }) \over i+1}
F(\alpha ,-\alpha ,j+1; {r^2 \over r^2 -1}) \bigr) \ .}}
The notation $(a)_{b}$ indicates Pochhammer symbols. From the
point of view of the scattering amplitude, an important property
of the kernel $K$ is that it contains the information that is
missed in the naive analytic continuation of the previous results.

As a consistency check, we can examine the behavior of \amplp\ as
$r \to 0$.  From \amplf\ we can see that the Toeplitz determinants
reduce to 1 as $r\rightarrow 0$.  In addition, from \amplj\
$Z\rightarrow 1$ as $r \to 1$, so we expect that $\det (1-K)
\rightarrow 1$ as well.  In the limit  the kernel factorizes:
\eqn\amplr{K(l_i,l_j) \rightarrow r^2\cdot (-\alpha )_{l_i+1}r^{l_i}
(-\alpha )_{l_j+1}r^{l_j} \ .}
As $r \to 0$ this  of course implies that $\det ( 1 - K) \to 1$.
In fact, a stronger statement is true -- $\det K$ vanishes to
leading order in $r$:
\eqn\ampls{\eqalign{\det [K(l_i,l_j)]^m_{i,j=1} &=
r^{2m} \prod^m_{i=1} (-\alpha )_{l_i}r^{l_i} \cr
& \times \sum_{\pi \in S_m} (-1)^{\pi} (-\alpha )_{l_{\pi (1)}}
r^{l_{\pi (1)}} \cdots (-\alpha )_{l_{\pi (m)}}
r^{l_{\pi (m)}} = 0
 \ . } }
Because of the this the $\det (1 - K)$ factor has no effect on the
pole structure of the amplitude at $r=0$,  in agreement with what
we expect from the OPE of the tachyon vertex operators.

We already know that in the $r \to 1$ limit the Toeplitz
determinant reduces to the form \ampleb \ and the bulk-bulk
amplitude reduces to the bulk-boundary amplitude.  Since  the
expression $Z$ \amplj\ is singular as $r \to 1$, there must be a
compensating singularity in $\det(1-K)$.  This is more challenging
to isolate.   It would be very interesting to understand the
properties of the determinant better since many aspects of the
physics are clearly stored in it.

\subsec{What we learn}

We have given a prescription for defining the two point scattering
amplitude for decaying branes at general momenta, and demonstrated
that the usual tactic of naively continuing from imaginary integer
momenta is  a large N approximation that misses important
contributions. The naive continuation fails because, remarkably,
the general amplitude localizes to the large N contribution at
these special momenta.  Since we are using analytic continuation
in our methods it is also possible that non-analytic contributions such
as zero-momentum delta functions are missed.  These are also
essential to understand, because they make contributions to
important quantities such as the stress tensor.  In any case, it
is clear that understanding the properties of Toeplitz
determinants emerging from U(N) random matrix theory will lead to
progress in the study of decaying branes and vice versa.

A useful way of making  progress with the bulk-boundary amplitude
in Sec.~4 was to do the integral over $x^0$ first, thus avoiding
the need to carry out a perturbative sum over $N$ as in \ampb.
Instead, the result then depends on a single $U(N)$ integral with
$N$ being related to the total energy in the process.  Such an
approach might be useful here since for large energies the simple
expression \amplj\ might suffice.

The general amplitude prior to integrating over the time $x^0$ is:
\eqn\ampbagain{
F(0,\omega_1,r,\omega_2;\mu)  =
\sum_{N=0}^\infty e^{-N \mu} \,  J_N(\omega_2,r)
~~~;~~~  \mu = -x^0 -\ln(-2\pi\lambda) \, .
}
This  has the form of a computation in a grand canonical ensemble
of $U(N)$ matrix models.  At very early times ($x^0 \to -\infty$)
the expression is dominated with smallest matrices.  As $x^0$
increases,  larger and larger matrices begin to play an important
role.  As $x^0$ increases beyond $-\ln(2\pi\lambda)$ infinitely
big matrices dominate the result and the formal power series fails
to converge.   To find the contribution to the amplitude from such
late times, we must analytically continue the fully summed
expression from convergent region $x^0 < -\ln(2\pi\lambda)$.  In
view of the many relationships between large U(N) matrices and
closed strings, as well as recent developments in 2d string theory
(\McGreevyKB, \KlebanovKM, \DouglasUP, \SenIV), it is tempting to conclude that as the brane decays
and spacetime becomes more closed-string-like in character, large
U(N) matrices emerge from the dynamics of the disappearing open
strings to describe the physical degrees of freedom of spacetime.

\newsec{Discussion}

Several  interesting questions in the physics of decaying branes,
including the nature of the final ``tachyon matter'' state and the
possibility of defining a decoupling limit leading to timelike
holography, require a deeper understanding of how strings scatter
from such unstable objects.   This has been a difficult problem to
solve because the decay  involves a non-trivial boundary
interaction on the open string worldsheet.   The standard
technique to calculate amplitudes in this system is to first
perform the necessary integrals at a special discrete set of
momenta and then optimisitically ``analytically continue'' the
resulting formal expressions.     Of course this procedure is not
strictly well-defined -- an analytic function cannot be defined
from its values at a discrete set of arguments without further
information, such as either the behavior at infinity or
consistency conditions derived from symmetry arguments.   However,
in the absence of further information this is the best we can do.
In this paper we have made progress in arriving at a
better-defined prescription for computing amplitudes.

First we showed that the timelike part of general amplitudes could
be written in terms of correlators in $U(N)$ matrix models. (While
we only studied tachyon scattering, vertex operators with more
oscillators could be treated in a similar manner.)   In a
perturbative approach, a grand canonical ensemble of matrix models
appeared, while integrating over the zero mode led to a single
matrix expectation value.  A key point is that the resulting
open-closed 2-point amplitude can be exactly computed in an open
subset of the complex momentum plane using Selberg's integral.
Analytic continuation from this region to real momenta is
reliable.  Indeed, we were able to compute the open-closed
amplitude in closed form in terms of zeta functions, and we
derived the asymptotics of closed string emission when the initial
condition for the decay is perturbed by an open string operator.  In particular we found that for extended branes the emission of heavy closed strings could be enhanced or suppressed by the presence of the additional vertex operator, with interesting consequences for the structure of the ``tachyon matter'' state to which the brane decays.
We likewise showed that the closed-closed scattering amplitude can
be expressed exactly in terms of  certain Toeplitz determinants of
hypergeometric functions.   The results obtained by naive analytic
continuation from integer momenta amount to the leading large $N$
approximation (due to Szeg\"{o}) of these determinants. In time
dependent quantum field theories, different choices of analytic
continuation in amplitudes are often related to different choices
of vacuum state.  We did not address the issue of the vacuum in detail here, but it would interesting to relate choices of analytic continuation to choices in the definition of the boundary state for the decaying brane.  (See \LambertZR, \MaloneyCK, \SchomerusVV, \FredenhagenUT, for some relevant discussion.)

In this paper we have focussed on the bulk-bulk and bulk-boundary
two point functions, and it is an important problem to
study more general correlation functions.  For example, it would
be interesting to study the boundary-boundary correlation function
in depth using the ideas developed here.   One would like to make
contact with the Liouville based approach in, e.g., (\GutperleXF, \StromingerFN, \SchomerusVV, \FredenhagenUT, \Volkertoappear).

 In addition  to the techniques described in the main text there are other powerful tools that can be brought to bear on the subject of scattering from decaying branes.  In  Appendix C we show, by extending old techniques of Douglas \Douglas, that there is a free fermion formulation of scattering from decaying branes.  This is related to the free fermion techniques  that can be used to study
 decaying branes  in 2d string theory (\McGreevyKB, \KlebanovKM, \DouglasUP, \SenIV).  Appendix D points out the relation between our
 scattering problem and a grand canonical ensemble for a classical Coulomb gas.
It would be very interesting to
apply these methods, and the techniques we have developed in
the main text, to a study of tachyon matter and to explore potential decoupling limits
leading to timelike holography, perhaps in the context of recently discovered non-singular
spacetime backgrounds representing decaying branes (\JonesRG, \Wang, \TasinatoDY,\LuIV).

\bigskip

\noindent {\bf Acknowledgments: } We are grateful to  
Jan de Boer, Neil Constable, Eric D'Hoker, Michael Gutperle, Antti Kupiainen, and Volker
Schomerus for useful conversations. Some of this work was carried
out during the ``Time and String Theory" workshop at the Aspen
Center for Physics. V.B. was supported by the DOE under grant
DE-FG02-95ER40893 and by the NSF under grant PHY-0331728.  E.K-V.
was supported in part by the Academy of Finland.  P.K. was supported in part by NSF grant PHYT-0099590.  A.N. was supported by Stiching FOM.

\appendix{A}{Special momenta and summing over $N$ in the bulk-boundary amplitude}

 At late (early) times $x^0$, the
chemical potential $\mu$ \jac, is large and negative (positive).
Thus  at late times, the series expression \nybb\  combined with
the Selberg result \ec\ for the bulk-boundary amplitude does not
converge. To get an explicit function of $x^0$  we must sum \nybb\
at early times and then analytically continue. Such a procedure is
rather delicate -- terms that are small at early times may
dominate after the analytic continuation to late times, so the sum
needs to be done exactly, or at least with attention paid to the
pieces that dominate late time behavior.   Progress can be made
for special momenta.

We can write the amplitude \nybb\ as
\eqn\a{ F(0,\omega_1,1,\omega_2;\mu) = 1+ \sum_{N=1}^\infty x^N \,
\prod_{j=1}^N {\Gamma(j) \Gamma(j+2i\omega_2) \over
(\Gamma(j+i\omega_2))^2} = 1+\sum_{N=1}^\infty    x^N \,
c(N,\omega_2) ~,}
with   $x = e^{-\mu}=-2\pi \lambda e^{x^0}$.   The coefficients
obey
\eqn\eh{ {c({N+1},\omega_2)\over c(N,\omega_2)}
= {\Gamma(N+1) \Gamma(N+1+2i\omega_2)\over
\left(\Gamma(N+1+i\omega_2)\right)^2} \, .}
A series of this form is, by definition, a hypergeometric function
if the ratio \eh\ is a rational function of $N$.
In our case, this happens if $i\omega_2$ is a positive
integer.
In fact, when $i\omega_2 = m$ is a non-negative integer a little algebra
shows that $c(N,-im)$ is a polynomial in $N$:
\eqn\d{ c(N,-im)=\prod_{j=1}^N {\Gamma(j) \Gamma(j+2m) \over
(\Gamma(j+m))^2} = \left(1+{N\over m}\right)^m \prod_{k=1}^{m-1}
\left[(1+{N\over k})(1+{N \over 2m-k})\right]^k.}
Furthermore $c(0,-im)=1$.

Setting  $i\omega_2=m$ the amplitude becomes $ F(1,-im;\mu)=  1 +
\sum_{N=1}^\infty x^N \, c(N,-im) $ which we can be written
\eqn\c{ F(0,\omega_1,1,\omega_2;\mu)= 1+ c(x{d \over dx},
-im)\sum_{N=1}^\infty  x^N= 1+ c(x{d \over dx},-im){x \over 1-x}.}
We thus arrive at
\eqn\f{ F(0,\omega_1,1,-im;\mu) =1+ \left(1+{x{d\over dx}\over
m}\right)^m  \prod_{k=1}^{m-1} \left[(1+{x{d\over dx}\over
k})(1+{x{d\over dx} \over 2m-k})\right]^k{x \over 1-x}.}
The simplest cases are
\eqn\g{\eqalign{  F(0,\omega_1,0,0;\mu) &= {1 \over 1-x} ~,\cr
F(0,\omega_1,1,-i;\mu)&=  {1 \over (1-x)^2} ~,\cr
F(0,\omega_1,1,-2i;\mu)&= -{1 \over 3} {x^3 -5x^2+x-3\over
(1-x)^5}~.}}

Note that the late time region is given by $\lim_{x_0\rightarrow
\infty} x = -\infty$.   It follows from \f\ that
\eqn\gaa{\lim_{x_0\rightarrow \infty}F(0,\omega_1,1,-im;\mu) =0.}
So we find that for these special values of momenta the
bulk-boundary amplitude vanishes at late time.

To summarize, \f\ combined with \nyac\ gives the result for the bulk-boundary two-point function at special momenta.

\bigskip

\appendix{B}{Heine's Identity}

We consider integrals of the form
\eqn\inta{I_n= \int \prod_{i=1}^{n} {dt_i \over 2\pi}
\prod^n_{i=1} f(t_i) \prod_{i <j} | e^{it_i}-e^{it_j}|^2 \ ,}
containing a $2\pi-$periodic function $f(t)$. In the Random matrix
literature this corresponds to an expectation value of a function
$\prod_{i=1}^{n}f(t_i)$ with respect to the circular unitary
ensemble \refs{\selbergs}. In our case, $f(t)=|1-ze^{-it}|^{2ik}$.
A well known identity due to  Heine \Heine, states that the expectation value of a
periodic function $f(t)$ with respect to the circular unitary
ensemble is equal to the Toeplitz determinant of the Fourier
coefficients $\hat{f}_k=\int f(t)
\exp (-ikt) dt$,
\eqn\intaa{{I_n \over n!} =D_n[f]=\det (\hat{f}_{(j-k)})_{1\leq
j,k \leq n} \ . }
A Toeplitz determinant is a determinant of a
(Toeplitz) matrix where all entries are equal along diagonals.
For convenience, we review the proof of Heine's identity \Heine.

First, note that the $\prod_{i <j} | e^{it_i}-e^{it_j}|^2$ term
is equal to $\Delta \bar{\Delta}$ where $\Delta$ is the Vandermonde
determinant $\Delta=\det\{e^{i(k)t_l}\}_{k,l=1, \cdots n}$. The integral
can be written as:
\eqn\intb{I_n= \int \prod_{i=1}^{n} {dt_i \over 2\pi}
\prod^n_{i=1} f(t_i) \Delta \bar{\Delta} \ . }
Now expand the determinants out
explicitly,  $\Delta=\sum_{\Pi}(-1)^{\Pi}e^{i\sum_{l=1}^{n}\Pi(l)t_l}$,
where $\Pi$ is a permutation of integers $1 \cdots n$.
Hence,
\eqn\intc{\Delta \bar{\Delta}
=\sum_{\Pi_1,\Pi_2}\prod_{l=0}^{n-1} (-1)^{\Pi_1+\Pi_2} e^{i
\Bigl( (\Pi_1(l)-\Pi_2(l))t_{l}\Bigr)} ~.}
Inserting this in \intb,
we obtain
\eqn\intd{I_n=\sum_{\Pi_1, \Pi_2}(-1)^{\Pi_1+\Pi_2}
\prod_{l=1}^{n}\bigg\{\int {dt_l \over 2 \pi}
f(t_l)~e^{i [
(\Pi_1(l)-\Pi_2(l))t_{l}]}\bigg\}~.}
Let $\nu_l=\Pi_1(l)-\Pi_2(l)$.  Then
\eqn\inte{I_n=\sum_{\Pi_1, \Pi_2}(-1)^{\Pi_1+\Pi_2}
\prod_{l=1}^{n}\bigg\{\int {dt_l \over 2 \pi} f(t_l)~e^{i\nu_l t_l}\bigg\}~.}
We can fix the first permutation $\Pi_1={1,2,3,\cdots,n}$ and
multiply the result by $n!$.
This simplifies the above expression:
\eqn\intg{\eqalign{I_n &=n! \sum_{ \Pi}(-1)^{\Pi}\prod_{l=1}^{n}
\bigg\{\hat{f}_{(l-\Pi(l))}\bigg\}\cr &=n! \det
(\hat{f}_{(l-m)})_{1\leq l,m\leq n}\ . }}
The determinant in the final expression is a Toeplitz
determinant of the Fourier coefficients of $f(t)$.

\bigskip

\appendix{C}{Fermionization}
In this appendix, we outline a relation between the closed string
scattering  amplitudes and correlation functions in a theory of
free fermions on a circle which gives an interesting reformulation
of the scattering problem. The basic idea is due to Douglas
\Douglas\ who showed the relation between U(N) group theory and a
theory of $N$ free fermions on a circle. As we saw in section 3,
the scattering  amplitudes can be written as a sum of correlators
in a grand canonical ensemble of unitary matrix models, with time
setting the fugacity: \eqn\fxz{F(x_0,z,\omega_c)=\sum_{N=1}^\infty
(-x)^N \int dM_{U(N)} | \det(1-zM)|^{2i\omega_c}~,} where $x=2\pi \lambda
e^{x_0}$. Using techniques developed in \Douglas, we can show that
this can be written as \eqn\fxzfer{ F(x,z,E_2)=\sum_{N=1}^\infty
(-x)^N {}\langle N | e^{ \int d\theta
\Psi^\dagger(\theta)f(\theta)\Psi(\theta)}| N \rangle ~,} where
$f(\theta)=-\mu \ln(1+r^2-2r\cos\theta)$ and $|N \rangle$ is the
$N$-fermion vacuum state satisfying \eqn\ffg{\eqalign{
B_n|N\rangle&=0, ~~~~|n|>{{N-1}\over 2}~,\cr
B^\dagger_{-n}|N\rangle&=0, ~~~~|n|\leq{N-1\over 2}~.}} Here, we
have expanded the fermionic fields $\Psi(\theta)$ and
$\Psi^\dagger(\theta)$ in terms of creation and annihilation
operators as \eqn\ffh{\Psi(\theta)=\sum_{n\in Z} B_n e^{in\theta},
~~~~~~~ \Psi^\dagger(\theta)=\sum_{n\in Z} B^\dagger_{-n}
e^{-in\theta}~.} The expression in \fxzfer\ is still quite
cumbersome because we have to sum over expectation values in the
$N$ fermion ground state. To obtain a more useful expression, we
will  add a factor of $e^{-\beta H}$ and sum over all states in
the Hilbert space. Then, by projecting onto the $N$ fermion sector
and taking the $\beta \rightarrow \infty$ limit correctly
reproduces the sum over $N$ fermion ground states. Thus the sum in
\fxzfer\ can be written as the $\beta \rightarrow \infty$ limit of
the following partition function: \eqn\partition{
Z(\beta)=\Tr\Bigl( e^{-\beta H} e^{ \int d\theta
\Psi^\dagger(\theta)f(\theta)\Psi(\theta)} (-x)^N P_N\Bigr)~, }
where $P_N$ is the projection operator onto the $N$ particle
sector: \eqn\projector{P_N=\int\!d\alpha\, e^{i \alpha
(\hat{N}-N)}~.} Here, $\hat{N}=\int d\theta \Psi^\dagger \Psi$
counts the number of fermions. The Hamiltonian  $H$  in the $N$
fermion sector is given by \eqn\ffe{H\equiv H_0 + E_0={1 \over
2}\int d\theta~ {d \Psi^\dagger \over d \theta} {d \Psi \over
d\theta} -{N(N^2-1)\over 12}~,} where we have defined $H_0={1 \over
2}\int d\theta~ {d \Psi^\dagger \over d \theta} {d \Psi \over
d\theta} $ and $E_0=-{N(N^2-1)\over 12}$. Notice that the $N$
fermion ground state $| N \rangle $ has zero energy. Using these
definitions we can write \partition\ as \eqn\partitiona{
Z(\beta)=\int d\alpha \Tr\Bigl( e^{-\beta H_0} e^{\int d\theta
\Psi^\dagger(\theta)(f(\theta) +i\alpha) \Psi(\theta)}\Bigr)\sum_N
e^{-i(\alpha+\pi)  N+\beta E_0 +\ln x}~. }

To proceed further, we notice that the trace in this expression
can be computed by  path integral methods.:
\eqn\pathinto{\Tr(e^{-\beta H} O )= \int_{\Psi(0)=\Psi(\beta)}
{\cal D}\Psi^\dagger {\cal D}\Psi e^{-S} O~.} The path integral
naturally computes the expectation values of operators which are
normal ordered, which here means that all the $\Psi^\dagger$'s are
pushed to the left. Thus to compute the trace \eqn\trace{\Tr\Bigl(
e^{-\beta H_0} e^{\int d\theta \Psi^\dagger(\theta)(f(\theta)
+i\alpha) \Psi(\theta)}\Bigr)} from a path integral, we  write the
operator $e^{\int d\theta \Psi^\dagger(\theta)(f(\theta) +i\alpha)
\Psi(\theta)}$ as a normal ordered operator: \eqn\normal{e^{\int
d\theta \Psi^\dagger(\theta)(f(\theta) +i\alpha) \Psi(\theta)}= :
e^{\int d\theta \Psi^\dagger(\theta)e^{(f(\theta) +i\alpha)}
\Psi(\theta)}:~.} Then, the trace can be written as a path integral
expression as follows: \eqn\pathint{\eqalign{\Tr\Bigl( e^{-\beta
H_0} e^{\int d\theta \Psi^\dagger(\theta)(f(\theta) +i\alpha)
\Psi(\theta)}\Bigr) = &\int
 {\cal D}\Psi^\dagger{\cal D} \Psi e^{\int d\theta dt \Psi^\dagger  \Bigl(
i{\partial \over \partial t}+{1 \over 2}{\partial^2 \over \partial
\theta^2}+e^{(f(\theta)+i\alpha)}\delta(t)\Bigr)\Psi} \cr = &
\det\Bigl( i{\partial \over \partial t}+{1 \over 2}{\partial^2
\over \partial \theta^2}+e^{i\alpha}
(1+r^2-2r\cos\theta)^{-\mu}\delta(t) \Bigr)}} where the path
integral is performed over periodic $\Psi$'s satisfying
${\Psi(0,\theta)=\Psi(\beta,\theta)}  $. This determinant is still
non-trivial to compute and here we do not push it any further. If this
determinant is computed, then the $\beta \rightarrow
\infty$ limit of $Z(\beta)$ in \partitiona\ can give the amplitude
\fxz.

\appendix{D}{Relation to a grand canonical ensemble of a classical gas}

It is interesting to note that the series \ampb\ can also be
rewritten as the grand canonical partition function of a
Coulomb gas of point charges in two dimensions,
confined in a circle and interacting with two external point
charges via repulsive Coulomb interactions \BakerForresterPearce .
The point charges
on the circle interact via a two body potential
\eqn\amplt{V(t_i,t_j)=-\ln |e^{it_i}-e^{it_j}| \ .}
The vertex operators at $z_1=0$ and $z_2=r$ correspond to two
external point charges at $1/z^*_1=\infty$
and $1/z^*_2 =1/r$ of strength $-\alpha$, acting on the charges
on the circle by the potential
\eqn\amplu{V(t_j)=\alpha \sum_{a=1,2}\ln |1-z_ae^{it_j}|
=\alpha \ln |1-re^{it_j}| \ .}
Now, we can introduce a temperature $1/\beta = 1/2$, and the
coefficients $J_n$ \amplc\ can be interpreted as the partition
function of a gas of $n$ charges interacting with the two external
ones at temperature $T=1/2$:
\eqn\ampluu{Z_N(T)=J_N = \prod_a \int {dt_a \over 2\pi}~e^{-\beta
(\sum^n_{a=1} V(t_a) + \sum_{a<b} V(t_a,t_b))} \ . }
If we formally introduce a fugacity $z=e^\mu = -2\pi \lambda
e^{x^0}$, the infinite series \ampb\ becomes the grand canonical
partition function:
\eqn\amplv{Q(T,z) = F(r,\alpha ; x^0) = \sum^{\infty}_{N=0}
z^N Z_N(T) \ .}
Note that for $\lambda >0$, the fugacity has
a ``wrong" negative sign. Further, eventually we will want $\alpha =
i\omega_2$, so that the strength of the charge at $z_2=1/r$
becomes imaginary.

\listrefs

\end